\documentclass[%
reprint,
superscriptaddress,
amsmath,amssymb,
aps,
pre,
]{revtex4-1}

\usepackage{dcolumn}
\usepackage{float}
\usepackage{bm}
\usepackage{comment} 
\usepackage{hyperref}
\usepackage{amssymb,amsmath,graphicx,array,color,amsthm}
\usepackage{caption}
\usepackage[skip=-5pt]{subcaption}
\usepackage{amscd}
\usepackage{multirow}

\def\bfk{{\bf k}}
\def\bfx{{\bf x}}
\def \rmd {\mathrm{d}}
\def \St {\mathrm{St}}
\def \cs {c_\text{s}}
\def \nB {n^\text{B}}

\begin{document}
\preprint{AIP/123-QED}

\title{Turbulence and far-from-equilibrium equation of state of Bogoliubov waves in Bose-Einstein Condensates}%

\author{Ying Zhu}
\email{Corresponding author: zhuying0325@gmail.com}
\affiliation{Universit\'{e} C\^{o}te d'Azur, CNRS, Institut de Physique de Nice {\color{black}(INPHYNI)}, 17 rue Julien Lauprêtre, 06200 Nice, France}
\author{Giorgio Krstulovic}
\email{Giorgio.KRSTULOVIC@univ-cotedazur.fr}
\affiliation{Universit\'{e} C\^{o}te d'Azur, CNRS, Institut de Physique de Nice {\color{black}(INPHYNI)}, 17 rue Julien Lauprêtre, 06200 Nice, France}
\affiliation{Universit\'{e} C\^{o}te d'Azur, Observatoire de la C\^{o}te d'Azur, CNRS, Laboratoire Lagrange, Boulevard de l'Observatoire CS 34229 -- F 06304 Nice Cedex 4, France}
\author{Sergey Nazarenko}
\email{Sergey.NAZARENKO@univ-cotedazur.fr}
\affiliation{Universit\'{e} C\^{o}te d'Azur, CNRS, Institut de Physique de Nice {\color{black}(INPHYNI)}, 17 rue Julien Lauprêtre, 06200  Nice, France}

\begin{abstract}
Bogoliubov waves are fundamental excitations of Bose-Einstein Condensates (BECs). They emerge from a perturbed ground state and interact nonlinearly, triggering turbulent cascades. Here, we study turbulent BECs theoretically and numerically using the 3D Gross-Pitaevskii model and its associated wave-kinetic equations. We derive a new Kolmogorov-like stationary spectrum for short Bogoliubov waves and find a complete analytical expression for the spectrum in the 
  long-wave acoustic regime. 
  We then use our predictions to explain the BEC equation of state reported by \textbf{Dogra et al. (Nature 620, 521, 2023)}, and to suggest new experimental settings.
\end{abstract}
\maketitle  

\section{Introduction}\label{intro}

Turbulence, in the broad sense, is perhaps one of the most striking manifestations of far-from-equilibrium physics in Nature. In classical fluids, it emerges when the forcing scale, at which energy is injected, is well-separated from the one at which it is dissipated by molecular viscosity. Energy is then carried along scales through a turbulent cascade driven by hydrodynamic interactions. An analogous process occurs in nonlinear wave systems, where cascades are driven by the interaction of waves having different wavelengths. When the wave nonlinearity is weak, the Wave Turbulence Theory (WTT) provides a comprehensive theoretical framework describing far-from-equilibrium steady and evolving states \cite{nazarenko2011wave}. The theory has been successful in describing, for instance, the dynamics of gravitational waves \cite{galtier2017turbulence}, Kelvin waves on superfluid vortices~\cite{Lvov_WeakTurbulenceKelvin_2010}, internal gravity waves in the ocean \cite{Caillol_KineticEquationsStationary_2000,Galtier_WeakInertialwaveTurbulence_2003}, Langmuir waves in plasmas \cite{Zakharov_CollapseLangmuirWaves_} and, more importantly for this work, matter waves in turbulent Bose-Einstein Condensates (BECs) \cite{dyachenko1992optical,zhu2023direct}.

One of the most interesting and important regimes in BEC turbulence is realized when weak waves propagate on a background of a strong coherent ground state $\Psi_0$, the condensate. 
These excitations are the Bogoliubov waves, 
whose frequency $\omega$ is given by the   dispersion relation,
\begin{equation} \label{eq:Bogodis}
    \omega_{k}\equiv\omega(k)=c_s k\sqrt{1+(k\xi)^2/2}\,,
\end{equation}
where $k=|\bfk|$ is the modulus of the wave vector $\bfk$, $c_s$ is the speed of sound, and $\xi$ is the healing length of the BEC at which dispersive effects become important.
The wave field $\Psi(\bm{x},t)$ can be viewed as a ground condensate with superimposed perturbations that $\Psi(\bm{x,}t)= \Psi_0(\bm{x,t})+\delta\Psi(\bm{x,t})$. Bogoliubov waves emerge when the condensate amplitude  $|\Psi_0|$ is significantly larger than the perturbations $|\delta\Psi|$, leading to quadratic nonlinearity and 3-wave interactions.

Over the last few years, outstanding progress has been achieved in realizing turbulent BECs experimentally, and they have become an excellent platform for studying far-from-equilibrium physics.
Turbulent states are driven by external forcing and dissipation, which results in an energy flux $P_0$ through scales (which equals the energy injection and dissipation rates). Such fluxes thus play a role equivalent to that of temperature in equilibrium thermal states. It is then natural to ask if there is a universal far-from-equilibrium equation of state (EoS) that relates energy flux and internal observables, such as energy or particle momentum distribution. The clean and well-controlled turbulent BEC experiments available today \cite{dogra2023universal}, together with recent numerical simulations of the Gross-Pitaevskii equation \cite{martirosyan2024_EoS_GP}, have led to revealing the existence of such EoS. They have shown that the amplitude of the wave-action spectrum scales with the flux as $P_0^{1/3}$, followed by a scaling $P_0^{2/3}$-scaling for large flux. Whereas the $P_0^{1/3}$-scaling can be understood within the WTT invoking 4-wave interactions, the second scaling-law lacks explanation. 

In the first part of the results, we obtain new theoretical predictions for 3D BEC turbulence dominated by interacting Bogoliubov waves, which are then confirmed by our numerical simulations. 
Among the most significant outcomes of our work are the following stationary energy spectra of Kolmogorov-Zakharov (KZ) type, which include the universal constants, for the acoustic and short wave limits, respectively:
\begin{align}
    E(k) &= C_1 c_s^{1/2} P_0^{1/2} k^{-3/2}, && \text{for } k\xi \ll 1, \label{eq:SZspectrum} \\
    E(k) &= C_2 c_s^{1/2} \xi^{5/2} P_0^{1/2} k, && \text{for } k\xi \gg 1, \label{eq:KZshort}
\end{align}
where the universal constants are derived as $C_1=1/{\sqrt{3\pi(\pi+4\ln{2}-1)}}$ and $C_2={2^{3/4}}/{\sqrt{\pi(\pi-4\ln{2})}}$.
As an application of our theory, in the second part of the results, we provide a plausible theoretical explanation of the second scaling law of the EoS observed experimentally in \cite{dogra2023universal}. 
This experiment used a cylinder-shaped box trap that was shaken to excite turbulence and which had a "finite height" thereby providing a dissipation (escape) of high-momentum particles. We re-analyze the experimental data of \cite{dogra2023universal}, and show their agreement with our prediction
of scaling $P_0^{1/2}$ arising from interacting short-wave Bogoliubov modes.
Our findings are important for BECs in a broad context, and as they are universal and expected to be generically realizable in turbulent systems, their impact thus goes beyond the explanation of a particular experiment. 

\section{Results}\label{results}
\subsection{Wave Turbulence Predictions for Bogoliubov waves}
\subsubsection{Stationary spectra of Komolgrov-Zakharov type}
 The master model for describing BECs is the  Gross-Pitaevskii equation (GPE) for the complex scalar wave function $\Psi(\bfx,t)$, where $\bfx$ is the position in 3D physical space and $t$ is time. Defining $\Psi$ such that $|\Psi|^2$ is the mass density, the GPE, written in terms of $\xi$, $c_s$ and the unperturbed (uniform) condensate mass density $\rho_0$, reads
\begin{equation} \label{eq:GP}
    i \frac{\partial\Psi  }{\partial t}=\frac{c_s}{\sqrt{2}\xi}\Big [ -\xi^2\nabla^{2}  +\frac{|\Psi|^{2} }{\rho_0}- 1\Big ]\Psi +V(\bm{x},t)\Psi
   +i\,\mathcal{F} - i\,\mathcal{D}\,.
\end{equation}
We have also included an external potential trap $V(\bm x)$, a large-scale forcing $\mathcal{F}$ and a hyper-viscous dissipation term $\mathcal{D}$ acting at small scales. In BEC experiments,  dissipation is achieved synthetically by allowing high energy particles to escape from the trap thereby removing energy from the system. In our simulations we mimic this effect by the hyper-viscous term which effectively  acts on the high-$k$ particles only. Moreover, in experiments, energy injection is provided by shaking the confining trap. First, in our idealized simulations, we set $V\equiv0$, but we include an external forcing in order to obtain an non-equilibrium steady state. Then, we verify our predictions in a trapped BEC, with an oscillating trap (and $\mathcal{F}=0)$. 
The healing length and the speed of sound depend both on the physical properties of the BEC and on $\rho_0$. 

The GPE can be mapped to an effective compressible irrotational fluid flow via the so-called Madelung transformation, $\Psi(\bfx,t) = \sqrt{\rho(\bfx,t)}\exp[i\phi(\bfx,t)/\sqrt{2}c_s\xi]$ with $\rho(\bfx,t)$ and $\phi(\bfx,t)$ being the fluid mass density and the velocity potential, respectively. The Bogoliubov dispersion relation \eqref{eq:Bogodis} is obtained by linearizing the GPE, \eqref{eq:GP}, in the absence of the potential, forcing, and dissipation terms, considering perturbations over the ground state, $\delta \Psi = \Psi - \sqrt{\rho_0}$, and diagonalizing the linearized equations in Fourier space. 

In our theory, we consider scales much smaller than the size of the potential trap, which allows us to consider the large-box limit and $V(\bm x) \equiv 0$.
Then, for weakly nonlinear waves, the leading order interaction occurs between the modes satisfying 3-wave resonant conditions,
\begin{equation} \label{eq:resonance}
\omega_{k}=\omega_{k_1}+\omega_{k_2}\,,\qquad  \bfk=\bfk_1+\bfk_2 \,.
\end{equation}
In this limit, the WTT provides a wave-kinetic equation (WKE) describing the evolution of the wave-action spectrum $\nB_{\bfk} =  \nB({\bfk},t)$
which is the Bogoliubov quasi-particle momentum distribution defined in \eqref{eq:bognk} in Methods.
The general WKE for weakly nonlinear waves driven by 3-wave resonant interactions is
\cite{ZLF,nazarenko2011wave}
 \begin{equation}
 \frac {\partial \nB_{\bfk}}{\partial t}= \St_{\bfk},
 \end{equation}
with the wave-collision integral
  \begin{equation}\label{eq:WKE}
    \begin{split}
  &\St_{\bfk}= 2 \pi \int \left(\mathcal{R}^{\bfk}_{\bf{12}} -\mathcal{R}^{\bm{1}}_{\bm{k2}} -\mathcal{R}^{\bm{2}}_{\bm{k1}} \right) \rmd \bfk_1 \rmd \bfk_2\,, \\
 & \mathcal{R}^{\bfk}_{\bm{12}} =        
  \delta^{\bfk}_{\bm{12}}   \delta(\omega^{k}_{{12}})
|V^{k}_{{12}}|^2  \left[ 
     \nB_{  {\bfk}_1} \nB_{  {\bfk}_2} - \nB_{  {\bfk}} \nB_{  {\bfk}_1} - \nB_{  {\bfk}} \nB_{  {\bfk}_2} \right] \,, \\
&\delta^{\bfk}_{\bf{12}} = \delta( {\bfk}  - \bfk_1 - \bfk_2)\,,
 \\
&\delta(\omega^{k}_{{12}}) = \delta(\omega_{k} - \omega_{  {k}_1} - \omega_{  {k}_2}) \,,
\end{split}
\end{equation} 
 where 
 $V^{k}_{{12}} \equiv V({k},{k_1}, {k_2})$ is the 3-wave interaction amplitude.
%
For Bogoliubov waves
\begin{equation}\label{eq:VBogol}
\begin{split}
&V^1_{23}=V_0\sqrt{k_1k_2k_3}W^1_{23}\,,\quad V_0=\frac{3\sqrt{c_s}}{4\sqrt{2}} \,,\\
&W^1_{23}=\frac{1}{2\sqrt{\eta_1\eta_2\eta_3}} + \frac{\sqrt{\eta_1\eta_2\eta_3}}{6k_1k_2k_3}\left(\frac{k_1^3}{\eta_1}-\frac{k_2^3}{\eta_2}-\frac{k_3^3}{\eta_3}      \right) \,,    
\end{split}
\end{equation}
where $\eta_i\equiv\eta(k_i) = \sqrt{1+(k_i\xi)^2/2}$ for $i=1,2,3$. The Bogoliubov WKE
\eqref{eq:VBogol} was first derived in \cite{dyachenko1992optical} in a slightly different form with mistakes in the prefactor.
%
%
%
Finally, by assuming isotropy and performing an angle average, the WKE becomes  
\begin{align}\label{eq:wke_disp}
    &\frac{\partial \nB_k}{\partial t} =\St_k 
    = 4 \pi^2 V_0^2 \int\limits_{k_1\,,k_2 \ge 0}  \big(|W^k_{12}|^2\mathcal{T}^k_{12}\delta(\omega^k_{12}) \\
     &-|W^1_{k2}|^2\mathcal{T}^1_{k2}\delta(\omega^1_{k2})   
    -|W^2_{k1}|^2\mathcal{T}^2_{k1}\delta(\omega^2_{k1}) \big) 
   \,k_1^2 k_2^2\rmd k_1 \rmd k_2\,, \nonumber
\end{align}
where $\mathcal{T}^k_{12} = \nB_{k_1} \nB_{k_2} - \nB_k \nB_{k_1} - \nB_k \nB_{k_2}$ (Methods). 
Note that only resonant waves interact.


\textbf{Acoustic limit. ---}
In the large-scale limit $k_i\xi\to 0, \;$ 
we have $\eta_i\to 1$ and therefore $\omega_{k_i}=c_s k_i $ -- this is the acoustic-wave limit. Also,  on the resonant manifold \eqref{eq:resonance}, we have $W^1_{23}\to1$, and thus $V^1_{23}\to V_0\sqrt{k_1k_2k_3}$, consistently with the result in \cite{griffin2022energy}. The collisional integral of the WKE \eqref{eq:wke_disp} becomes as in the 3D acoustic WKE \cite{zakharov1970spectrum},
\begin{equation}\label{eq:wke_acous}
  \St_k= \frac{4 \pi^2 V_0^2}{ c_s} \!\!\!\!\!\!\! \int\limits_{\, k_1,k_2 \ge 0}  \!\!\!\!\!\! \left(\mathcal{T}^k_{12}\delta^k_{12}
  -\mathcal{T}^1_{k2}\delta^1_{k2} -\mathcal{T}^2_{k1}\delta^2_{k1} \right) \! k_1^2 k_2^2\rmd k_1 \rmd k_2,
\end{equation}
where $\delta^k_{12}=\delta(k-k_1-k_2)$. 

In \cite{zakharov1970spectrum}, Zakharov and Sagdeev obtained a Kolmogorov-type stationary solution for the energy spectrum of 3D acoustic waves:
$E(k)=4\pi k^2 \omega_k \nB_k=C_1\,c_s^{1/2}P_0^{1/2}k^{-3/2}$, the so-called  Zakharov-Sagdeev (ZS) spectrum. 
Here, the energy flux $P_0$ is constant because of the stationarity of the spectrum.
An approximation of the dimensionless constant $C_1$ was given in \cite{zakharov1970spectrum}, as well in a recent paper \cite{kochurin2024three}. We derive its exact analytical expression, and obtain the complete ZS spectrum,  \eqref{eq:SZspectrum} (Methods). 

Note that the acoustic WKE \eqref{eq:wke_acous}, is derived as an asymptotic limit of \eqref{eq:wke_disp}.
Unlike the 2D case where $St_k\propto \xi^{-1}$ \cite{griffin2022energy}, in 3D $\xi$ cancels out after angle averaging \cite{ZLF} leading to \eqref{eq:wke_acous}. However, for its validity, a small amount of dispersion, naturally present in BECs,  is still crucial in the system to regularize the wave dynamics and avoid singularities.
%

\textbf{Short-wave limit. ---}
In the small-scale limit $k\xi \to \infty$, we have $\eta_k\to k\xi/\sqrt{2}$ and $V^1_{23}\to 2^{7/4}V_0\xi^{-3/2}/3$. The collisional integral becomes
\begin{equation}\label{eq:wke_short}
\begin{split}
  \St_k& = \frac{32\sqrt{2}\pi^2 V_0^2}{9\xi^3k}   \int\limits_{k_1\,,k_2 \ge 0}  \big(\mathcal{T}^k_{12}\delta(\omega^k_{12}) 
   -\mathcal{T}^1_{k2}\delta(\omega^1_{k2})   \\
  &-\mathcal{T}^2_{k1}\delta(\omega^2_{k1}) \big) 
  \,k_1 k_2\rmd k_1 \rmd k_2\,.
\end{split}
\end{equation}
where now $\omega_k=c\xi k^2/\sqrt{2}$.
Note that this collision integral coincides with the one for the finite-$k$ spectral part $ {\nB_k}'$ obtained from the 4-wave kinetic equation
via substitution $\nB_k =N_0 \delta({\bf k}) +  {\nB_k}'$, $N_0=$~const; see e.g. \cite{during2009}.
 
To find the Kolmogorov-type spectrum in the short-wave limit, we apply the standard procedure provided by the WTT and seek a stationary power-law solution of \eqref{eq:wke_short}, $\nB_k=Ak^{-x}, \; A=$~const.  This leads to
$\St_k=64V_0^2\pi^2A^2(9\xi^4 c_s)^{-1} k^{1-2x}I(x)$, with the dimensionless collision integral
\begin{align}\label{eq:Ix}
&I(x)=\int\limits_{q_1\,,q_2\ge 0}
(q_1q_2)^{1-x}\Big(\left(1-q_1^x-q_2^x\right)\delta(1-q_1^2-q_2^2)  \nonumber\\
&-2\left(q_1^x-1-q_2^x\right)\delta(q_1^2-1-q_2^2) \Big)
\rmd q_1 \rmd q_2\,. 
\end{align}
The standard definition of the energy flux is given by $P(k)=-\int_0^k 4\pi\,q^2\omega(q)\St(q)\rmd q=\frac{128\sqrt{2}}{9}\pi^3V_0^2\xi^{-3}A^2\frac{I(x)}{2x-6}k^{6-2x}$. Constant energy flux, $P(k)\equiv P_0=$~const, implies $2x-6=0$, i.e. $x=3$ and $I(3)=0$. Using the L'H\^{o}pital rule, $\lim_{x\to3} I(x)/(x-3) = dI(x)/dx |_{x=3}$.  
We evaluate this expression and find the steady spectrum,  \eqref{eq:KZshort} (Methods).
This spectrum, as well as the constant $C_1$ in \eqref{eq:SZspectrum}, are new theoretical predictions. Note that the constants $C_1$ and $C_2$ are finite because the respective integrals defining the energy flux are convergent, a property called locality of the wave interactions (Methods). We will now test these predictions numerically via simulations of the GPE and the isotropic WKE. 

\subsubsection{Numerical Validation}
\begin{figure*}
\centering
  \includegraphics[width=\textwidth]{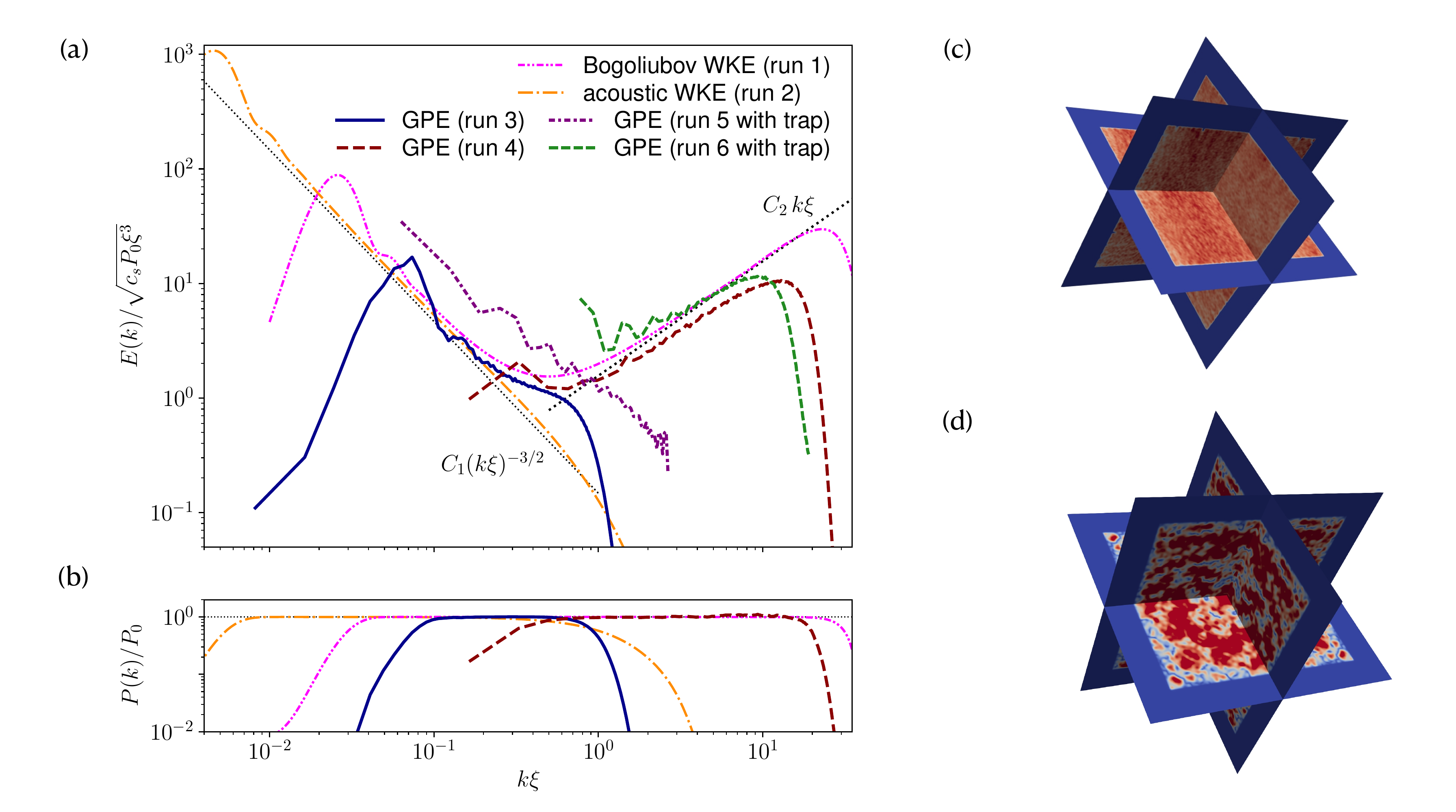}
  \caption{\label{fig:EkPk} 
  Numerical results obtained by simulating WKE and GPE.
  (a) Dimensionless energy spectra $E(k)/\sqrt{c_sP_0\xi^3}$ vs. dimensionless wave number $k\xi$ with  theoretical predictions \eqref{eq:SZspectrum} and \eqref{eq:KZshort} superimposed. Run 1 is performed using the WKE~\eqref{eq:wke_disp} and contains both the acoustic (long-wave) and short-wave regimes, whereas run 2 uses the $3D$ acoustic WKE \eqref{eq:wke_acous}. The GPE runs 3 and 4 are obtained for homogeneous cases with stochastic forcing, and runs 5 and 6 are obtained with a shaken potential trap. All the GPE runs are optimized such that the inertial range extends through the scales of interest.
  (b) Corresponding energy fluxes of run 1-4 normalized by their values measured in the inertial range. (c) Snapshot of density for GPE run 5 to verify the prediction in the acoustic limit. (d) Snapshot of density for GPE run 5 to verify the prediction in the short-wave limit. 
\vspace{-4ex}
}
\end{figure*}

 First, we numerically simulate the forced-dissipated GPE, \eqref{eq:GP}, setting $\rho_0=1$ and $c_s=1$, in a triply periodic cube with the side length $L=2\pi$, and $V(\bm x)\equiv 0$. To focus on different ranges with respect to the transitional wavenumber $k_\xi =1/\xi$, two high-resolution GPE simulations with $\xi=1.25 \Delta x$ and $\xi=15 \Delta x$ ($\Delta x$ is the length of the grid spacing) are performed.
Such choices allow us to study the acoustic and the short-wave limits separately.
We set the initial data with uniform condensate with $|\Psi(0)|^2=\rho_0$, a stochastic forcing then adds the wave disturbances, and we
evolve the system until it reaches a steady state, balancing between the large-scale forcing and the small-scale dissipation (Methods). We
then perform averages over time.
Also, we perform two isotropic WKE simulations: one of the Bogoliubov WKE, \eqref{eq:wke_disp}, and the other of the purely acoustic WKE, \eqref{eq:wke_acous}, with extra forcing and dissipation terms similar to the ones used in the  GPE simulations (Methods).

The results of these runs in terms of the energy spectra and the fluxes superimposed with the respective theoretical predictions,  \eqref{eq:SZspectrum} and \eqref{eq:KZshort}, are presented in Fig.~\ref{fig:EkPk}.
The energy fluxes shown in Fig.~\ref{fig:EkPk}(b)  exhibit plateau ranges, with $P(k)=P_0=$~const, in which the spectra are stationary and both forcing and dissipation effects are negligible (the so-called inertial ranges). The measured values of $P_0$ are then used for normalizing the energy spectra in Fig~\ref{fig:EkPk}(a).
We can see that each of the respective inertial ranges, for both WKE (run 1 and run 2) and GPE (run 3 and run 4) simulations, are in excellent agreement with our theoretical predictions (both for the spectral slopes and the prefactor constants). The longest inertial range, spanning for about two decades, is realized in the acoustic WKE run (run 2) with scaling spectrum \eqref{eq:SZspectrum} confirmed.  The Bogoliubov WKE run (run 1) produced  two scaling ranges in the large-scale and the short-scale parts of the spectrum simultaneously (about a decade-wide each). The GPE runs with small and big values of $\xi$
produced one scaling each for $k\xi<1$ (run 3) and  $k\xi>1$ (run 4) . 
The scaling range in the GPE run 3 is the narrowest among all the simulations, but its spectrum agrees with the Bogoliubov WKE simulation. In particular, it exhibits a bump in the high-$k$ part of the spectrum close to the transitional wave number, $k\sim k_\xi$. A similar bump was previously observed in GPE simulations of \cite{Dispersive}, and it was called a "dispersive bottleneck" since it is related to an increased wave dispersion in this region of scales. Note that our predictions are well confirmed by the GPE simulations without any adjustable parameters.

In BEC turbulence experiments, the cloud is always confined by an external potential trap. To be more realistic, we perform two further GPE simulations in a box-shaped potential trap of size $L_{\rm trap}=1.5\pi$, with $\xi=1.25 \Delta x$ (run 5) and $\xi =10 \Delta x$ (run 6), namely $L_{\rm trap}=230.4\,\xi$ and $L_{trap}=28.8\,\xi$, respectively. We prepare ground states so that the initial density inside the trap is nearly constant with $|\Psi(0)|^2=\rho_0$. After that, we shake the trap in both the $x$ and $y$ directions to introduce energy into the system until the small-scale energy dissipation rate becomes equal to the injection rate. We adjust the shaking protocol and the dissipation parameters to keep the system in the Bogoliubov regime ensuring that at least 70\% of the particles are staying at $k\le k_{trap}=2\pi/L_{\rm trap}$. We measure the energy flux $P_0$ by calculating the dissipation rate of energy on small scales, in the same spirit as in \cite{dogra2023universal} (Methods). Fig.~\ref{fig:EkPk}(c) and (d) show the density snapshots for runs 5 and 6, respectively.
Although the presence of traps reduces the width of the inertial range, 
Fig.~\ref{fig:EkPk}(a) shows that the spectra obtained with the potential trap remain consistent with the WTT-predicted scalings in both regimes  (run 5 for $k\xi \sim 0.08-0.5$,  and run 6 for $k\xi \sim 2-8$).
On the other hand, in both cases, the interaction with the potential trap increases the effective spectrum constants: 
by a factor $\sim 4$ in the acoustic regime and $\sim 1.2$ for the short waves.
 
\subsection{Interpreting 
the experimental EoS }
\subsubsection{EoS predicted by the 4-wave and the 3-wave WT}
Almost all BEC experiments relevant to wave turbulence have been interpreted within the framework of the 4-wave interaction theory \cite{navon2016emergence, dogra2023universal, moreno2025observation}, 
which assumes that the wave perturbations $\delta\Psi$ greatly exceed the condensate amplitude $\Psi_0$,
leading to cubic nonlinearity. Instead, in this work we  employ the new short-wave solution \eqref{eq:KZshort} governed by 3-wave interactions to interpret the experimental results of Dogra et al. \cite{dogra2023universal} on the far-from-equilibrium EoS. 
Dogra et al. analyzed the particle spectra (occupation number spectra) $n_k$ (which has a difference of prefactor of $8\pi^3$ from the one we defined as \eqref{eq:ntildek} in Methods) measured in the direct cascade of BEC turbulence by fitting a scaling-law of $k^{-3 + \delta}$, with $\delta$ being an adjustable parameter.
Note that at small scales, our prediction \eqref{eq:KZshort} corresponds to a $k^{-3}$ wave-action spectrum (equivalent to the particle spectrum, in this case, up to a prefactor difference; see \eqref{eq:nkshift} in Methods), which coincides, up to the log-correction, with the 4-wave prediction found in \cite{zhu2023direct}. This coincidence makes both spectra good candidates for explaining the experimentally found EoS, assuming that they correspond to the short-wave range  such that  $k\xi>1$. 

The direct cascade solution of the 4-wave WKE found in  \cite{zhu2022testing}, recast in a dimensional form, reads
\begin{equation} \label{eq:EoS4}
n_k=n_{\rm 4w} \, k^{-3}\ln^{-\frac{1}{3}}\left(\frac{k}{k_{\rm f}}\right)\,, n_{\rm 4w} =C_{\rm 4w}\left(\frac{m^2\epsilon}{\hbar^3 a_s^2}\right)^{\frac{1}{3}}\, ,
\end{equation}
where $k_{\rm f}$ is the forcing wave number and $C_{\rm 4w}={C_d}{(4\pi)^{-\frac{2}{3}}}$, with $C_{\rm d}\approx 5.26\times10^{-2}$ being a universal constant. 
Here,
$m$ is the atom's mass, 
$\hbar$  is the reduced Planck constant, and $a_s$ is the $s$-wave length. The energy flux $\epsilon$ is defined as the energy  dissipated in the system per unit time per unit  volume.

In \cite{dogra2023universal}, the authors performed a series of experiments with various parameters, interaction strength $a_s$, forcing intensity, and mean particle density $N$
such that $N=\int 4\pi k^2  n_k \, \rmd k$ with the definition of $n_k$ as in this paper. They obtained the particle spectra amplitude $n^{\rm exp}$ by fitting the experimental spectra with $n_k = n^{\rm exp} k^{-3.2}$, and discovered that employing the dimensional variables $n^{\rm exp}$ and $\epsilon$ as two state variables results in a non-universal EoS for different $a_s$ and $N$. Furthermore, they empirically found that the dependence of the non-dimensional spectrum amplitude, $n_{\rm exp}/N$ versus the non-dimensional energy flux $\tilde \epsilon = m^2 \hbar^{-3} a_s^{-2} N^{-3}\epsilon$ collapses the experimental data into a universal EoS.
Theoretical prediction \eqref{eq:EoS4}  implies 
\begin{equation}
n_{\rm 4w}/N = C_{\rm 4w} \tilde \epsilon^{1/3}, 
\label{eq:EoS4n}
\end{equation}
which was confirmed in 
\cite{dogra2023universal} for data points with low values 
of $\tilde \epsilon$. However, at larger $\tilde \epsilon$,  \cite{dogra2023universal} reported a steeper EoS,  $n_{\rm 4w}/N \sim \tilde \epsilon^{\alpha}$ with $\alpha >1/3$. 
The authors of \cite{dogra2023universal} noted that  $\alpha \approx 2/3$ provide a good fit implying a flux-dependence like the one of the classical Kolmogorov hydrodynamic turbulence.
However, it is unclear how a state with the classical Kolmogorov turbulence could appear in this experiment, considering that it is unlikely that the cascade process is generated by vortices with hydrodynamic properties. Indeed, such regimes are usually attributed to states with polarised vortex tangles at scales greater than the inter-vortex distance~\cite{Polanco_VortexClusteringPolarisation_2021}, which do not seem to be relevant to the considered experiment. Also, as noted in \cite{dogra2023universal}, the Kolmogorov spectrum would imply a spectral exponent $-5/3$ that is very different from the observed exponent $\approx -3$.

Here, we suggest a plausible interpretation of the steeper EoS.
We propose that a stronger ground-state condensate component could be present for greater $\tilde \epsilon$ making the system switch to the 3-wave turbulence of short Bogoliubov waves, implying $\alpha \approx 1/2 >1/3$.

Recasting our prediction \eqref{eq:KZshort} in the same manner as in
\eqref{eq:EoS4}, we obtain
 (Methods)
\begin{equation} \label{eq:EoS3}
n_k=n_{\rm 3w} k^{-3}\,, \frac{n_{\rm 3w}}{N}=C_{\rm 3w}\left(\frac{m^2\epsilon}{\hbar^3 N^3 a_s^2}\right)^{\frac{1}{2}}\,,C_{\rm 3w}=\frac{C_2}{2^{17/4}\pi^2}\,.
\end{equation}
The dimensional $P_0$  in \eqref{eq:SZspectrum} is related to $\epsilon$ as $\epsilon = \rho_0 P_0 $, since $P_0$ complies with the standard definition of the energy flux as the energy dissipation rate per unit mass whereas 
$\epsilon$ used in \cite{dogra2023universal} is the energy dissipation rate per unit volume.
The EoS corresponding to \eqref{eq:EoS3} is
\begin{equation}   
n_{\rm 3w}/N = C_{\rm 3w} \tilde \epsilon^{1/2}.
\end{equation}
This  EoS looks similar to the  4-wave EoS \eqref{eq:EoS4n}: the both laws are of the form $n/N=C\,\tilde \epsilon^\alpha$. Hovewer, one should not forget the important difference between the 4-wave and the 3-wave case: the non-normalised 4-wave EoS given by 
\eqref{eq:EoS4} does not contain any $N$-dependence whereas the 3-wave law given in \eqref{eq:EoS3} does involve $N$.
\begin{figure*}[htbp]
\centering
\includegraphics[scale=0.88]{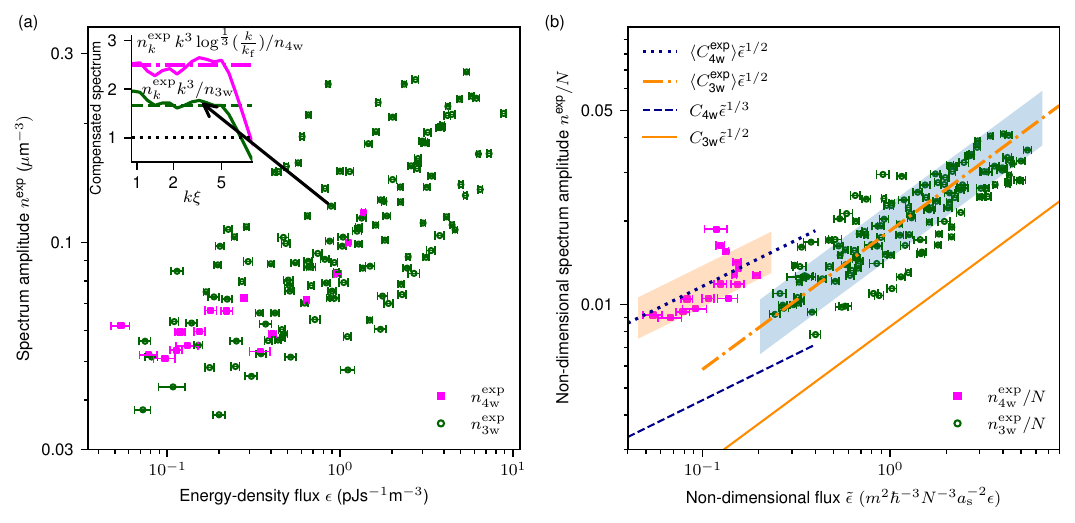}
\caption{\label{fig:EoS} Equation of State (EoS) regenerated from the experimental data of \cite{dogra2023universal}. 
(a) Particle spectrum amplitude $n^{\rm exp}$ .vs. energy flux $\epsilon$, where $n^{\rm exp}$ is taken either as $n^{\rm exp}_{\rm 4w}$ in magenta fitting by the 4-wave prediction \eqref{eq:EoS4} or $n^{\rm exp}_{3w}$ in green fitting by the 3-wave prediction \eqref{eq:EoS3}. The fitting procedure is indicated by the inset of (a): 
Each experimental spectrum $n^{\rm exp}_k$ is compensated by $k^3 \ln^{1/3}(\frac{k}{k_{\rm f}})$ and $k^3$, respectively. 
$n^{\rm exp}_{\rm 4w}$ and $n^{\rm exp}_{\rm 3w}$ are calculated by averaging the corresponding compensated spectrum 
on a plateau range. We take $n^{\rm exp}$, $n^{\rm exp}_{\rm 4w}$ or $n^{\rm exp}_{\rm 3w}$ depending on which is closer to its theoretical prediction, $n_{\rm 4w}$ or $n_{\rm 3w}$. (b) Non-dimensional particle spectrum amplitude $n^{\rm exp}/N$ .vs. non-dimensional flux $\tilde{\epsilon}$. Theoretical predictions $C_{\rm 4w}\tilde \epsilon^{1/3}$ and $C_{\rm 3w}\tilde \epsilon^{1/2}$ and the ones replacing $C_{\rm 4w}$ and $C_{\rm 3w}$ by the mean values of the respective constants obtained by fits of the experimental spectra, $\langle C_{\rm 4w}^{\rm exp} \rangle$ and $\langle C_{\rm 3w}^{\rm exp} \rangle$, are superimposed. The shaded bands around the latter two curves indicate one standard deviation, evaluated in logarithmic scale for the deviations of the data points relative to the fitted lines.
The horizontal error bars indicate the uncertainty associated with the flux, based on the uncertainty estimates provided in the original experimental data.
}
\end{figure*}

\subsubsection{Re-analysis of the experimental data}
The revised EoS based on the experimental data points of \cite{dogra2023universal}  is shown in Fig.~\ref{fig:EoS}.
Our procedure is different from the one of \cite{dogra2023universal}: for each experimentally obtained spectrum $n^\text{exp}_k$, 
we try our two theoretical predictions, namely
$n_k = n_{\rm 4w}^{\rm exp} k^{-3} \, \ln^{-\frac{1}{3}}\left(\frac{k}{k_{\rm f}}\right)$,
and $n_k = n_{\rm 3w}^{\rm exp} k^{-3}$, 
investigate the compensated spectra $n^\text{exp}_k k^3 \ln^{1/3}(\frac{k}{k_{\rm f}})$ and $n^\text{exp}_k k^3$, respectively, and compute $n^{\rm exp}_{\rm 4w}$ and $n^{\rm exp}_{\rm 3w}$ by averaging the corresponding compensated spectrum in their plateau ranges, as shown in the inset of Fig.~\ref{fig:EoS}(a) (see the Methods for an example where the 4-wave prediction is selected). 
We take the value of the spectrum amplitude, $n^{\rm exp}$, either $n^{\rm exp}_{\rm 4w}$ or $n^{\rm exp}_{\rm 3w}$, depending on which is closer to its corresponding prediction, $n_{\rm 4w}$ or $n_{\rm 3w}$. In the other words, we find the best fit for the compensated spectra in terms of the $L^1$-norm.
We plot the fitting spectrum amplitude $n^{\rm exp}$ vs. the flux $\epsilon$ in Fig.~\ref{fig:EoS}(a), magenta for the points taken with $n^{\rm exp}_{\rm 4w}$ and green for those taken with $n^{\rm exp}_{\rm 3w}$. 
The non-dimensional EoS is presented in Fig.~\ref{fig:EoS}(b), in which the points
split into two distinct clouds and collapse onto a universal EoS.
 Note that in \cite{dogra2023universal}, the authors reduced the spread of their EoS by introducing two extra phenomenological fitting parameters in addition to $n^{\rm exp}/N$ and $\tilde{\epsilon}$. Our fitting procedure is based only on comparison with our two theoretical predictions without any introducing any adjustable parameters.

Fig.~\ref{fig:EoS} (b) also shows
the theoretical 4-wave and 3-wave  EoS predictions in which the constants $C_{\rm 4w}$ and $C_{\rm 3w}$ are replaced by the mean values of the respective constants obtained by fits of the experimental spectra, $\langle C_{\rm 4w}^{\rm exp} \rangle$ and $\langle C_{\rm 3w}^{\rm exp} \rangle$. 
To provide a visual indication of the  fit quality, shaded bands are added around the two fitted curves; they correspond to one standard deviation in logarithmic scale of the data relative to the fitted lines.
We see that the theoretical predictions are consistent with the experimental data, with the 4-wave and 3-wave EoS in agreement with the low and high $\tilde \epsilon$ parts respectively.
The mean experimental constants are nearly twice the theoretically predicted values, $\langle C_{\rm 3w}^{\rm exp} \rangle \approx 2.2 C_{\rm 3w}$ and $\langle C_{\rm 4w}^{\rm exp} \rangle \approx 2.4 C_{\rm 4w} $.
The deviation of the experimentally measured constants from their theoretical predictions is similar 
of the same order as the spread ($L^\infty$-norm, $\Delta C^{\rm exp}= C^{\rm exp}_{\rm max}-C^{\rm exp}_{\rm min}$) of the EoS data points from the mean, $\Delta C^{\rm exp}_{\rm 4w} \approx 2.3 C_{\rm 3w}$ and $\Delta C^{\rm exp}_{\rm 4w} \approx 1.7 C_{\rm 4w}$.
Since there are no adjustable parameters in our theoretical results and fitting process, 
and theoretical predictions are asymptotical for $k\xi\gg 1$ (which in experiments are not satisfied in a strong sense, as $k\xi\approx 4$)
we conclude that our theoretical predictions reasonably agree with the experimental data.
Our numerical simulations with shaken traps, which more closely correspond to the experimental setup, show that the interaction with the trap tends to increase the effective constants of the stationary KZ-type spectra relative to the corresponding simulations without the trap. This trend is also consistent with our previous study of \cite{zhu2023direct}, and agrees to the observation that $\langle C_{\rm 4w}^{\rm exp}\rangle > C_{\rm 4w}$ and $\langle C_{\rm 3w}^{\rm exp} \rangle > C_{\rm 3w}$. A possible explanation is that, in the box-trap configuration, the zero boundary conditions replace the plane-wave eigenmodes of the homogeneous case by standing-wave, sine-like modes. 
This significantly reduces the number of active resonant modes and effectively decreases the prefactor of the WKE collision integral. As a consequence, for a given flux, this tends to produce a larger effective constant of the stationary KZ-type spectrum. This effect is expected to be more pronounced in smaller systems, where the available scale separation is insufficient.
We expect that in  larger traps, the experimental measurements would become closer to the universal limit and their measured constants would tend towards the predicted values. Nevertheless, further studies of the inhomogeneity induced by traps and respective systematic deviations in numerical (and experimental) flux measurements are desirable to better comprehend the finite size effects.
   

\section{Discussion}
In summary, in this paper we have obtained the Kolmogorov-type stationary spectrum of forced-dissipated turbulence of short Bogoliubov waves, \eqref{eq:KZshort}, and we revised the long-wave Kolmogorov-type spectrum including finding the exact prefactor constant $C_1$, see \eqref{eq:SZspectrum}.
We have validated these solutions, along with the combined spectrum including both long and short waves by numerical simulations of the GPE and the isotropic Bogoliubov WKE.
Our work relies on the turbulence universality hypothesis, stating that the inertial (non-dissipative) range spectrum is insensitive to specific forcing and dissipation mechanisms. Consequently, we adopted stochastic forcing and hyperviscous dissipation to maximize the inertial range, as is customary in turbulence simulations. Our prediction were also tested in more realistic settings, by using an oscillating external trap to force the system.
We used the newly found short-wave spectrum for interpreting the (previously unexplained)  experimental Equation of State of Ref.\cite{dogra2023universal}. To test our theory in future experiments,
the realizability of Bogoliubov-wave turbulence in both asymptotic limits should be further examined. For the short-wave regime, although the present comparison with \cite{dogra2023universal} already shows a reasonable agreement at $k\xi \sim 4$, a cleaner quantitative verification of the asymptotic prediction would benefit from experiments reaching larger values of $k\xi$.
For the long-wave acoustic regime, future verification is expected to require sufficiently large systems, with $L_{\rm trap}/\xi$ of order 100.
In the short-wave regime, one should additionally check: (i)
whether a strong ground state is present for strong forcing cases, 
because in the Gross--Pitaevskii model the relevant interaction depends on the background on which the fluctuations propagate: the three-wave description applies when $|\Psi_0| \gg |\delta\Psi|$, whereas the four-wave description applies in the opposite limit $|\Psi_0| \ll |\delta\Psi|$; (ii) whether, for these cases, the measured spectrum corresponds to waves shorter than the healing length and with amplitudes weaker than the ground state. 
A plausible route to the three-wave regime in the experiment is that, at large fluxes, the turbulent steady state is established before substantial long-time condensate depletion occurs, so that the short-wave fluctuations still propagate on a strong condensate background during the measurement window. At lower fluxes, by contrast, the steady state is established more slowly (see Fig.~2(a) of \cite{dogra2023universal}), and the longer waiting time may allow stronger condensate depletion, thereby favoring the four-wave regime. This interpretation, however, still requires direct experimental verification.
It should be emphasized that, within the WTT framework, the three-wave and four-wave regimes correspond to distinct asymptotic limits, namely $|\Psi_0| \gg |\delta\Psi|$
and $|\Psi_0| \ll |\delta\Psi|$, respectively. It is natural to expect that some points in a real experimental system do not fall into either of the two  ideal limits, but rather lie in an intermediate range where $|\Psi_0| \sim |\delta\Psi|$.
This may contribute to the remaining variance between our predictions and the experimental data. 
Therefore, our classification of the experimental points in Fig.~\ref{fig:EoS} is meant to identify the dominant scaling behavior and should be understood as a leading-order interpretation.
The transitional  regime corresponds to stronger waves whose statistics is no longer described by standard WTT. A future model is anticipated to describe short scales in this transitional regime, where the condensate amplitude and fluctuation amplitude are comparable.

Although the non-dispersive acoustic ZS spectrum is not the main topic of this work, it is an interesting side result that we would like to comment on.   
We find a way to regularize the WKE of non-dispersive acoustic waves for the 3D isotropic case, but get a different prefactor. Consequently, the ZS spectrum comes with a bigger constant (Methods).
In  recent work \cite{kochurin2024three}, the authors numerically obtained the ZS spectra in 3D acoustic wave systems, and different values of constant were reported for the dispersive and non-dispersive cases. 
The constant observed in \cite{kochurin2024three} for the non-dispersive acoustic system is smaller than the one for the dispersive system, unlike what we get here. Their paper offers an explanation based on a numerical observation of an anisotropy (i.e. presumably the isotropic spectrum is unstable with respect to anisotropic perturbations).  Note, however, that the model acoustic-wave equation studied in \cite{kochurin2024three} was different from the compressible Euler equation or GPE. Thus, the question about the acoustic wave spectrum arising in the non-dispersive Euler equation merits further study.

\section*{Methods}
\subsection*{Wave-kinetic equations and steady Kolmogorov-type spectrum}
The starting point to derive the 3-wave interaction amplitude $V^1_{23}$ in \eqref{eq:VBogol} is to rewrite the Hamiltonian of the second order $H_2$ into the canonical form of $H_2=\Sigma_{\bfk}\omega_\bfk a_\bfk a^*_\bfk$, then the wave-action spectrum is defined as 
\begin{equation}\label{eq:bognk}
\nB_\bfk=\left( \frac{L}{2\pi}\right)^3 \langle |a_\bfk|^2 \rangle\,,
\end{equation}
where the problem is considered in an $L$-periodic box.
As a consequence, one rewrites the Hamiltonian of third order into $H_3=\sum_{1,2,3}V^1_{2,3}\left( a_1a^*_2a^*_3 +\text{c.c} \right)\delta^1_{2,3}\,$ and gets the interaction amplitude. In the Supplementary Materials (SM), we provide two methods to derive $V^1_{23}$.
Just to mention one of the two methods here, we first  assume $\Psi(\bm{x},t)=\Psi_0(1+\psi(\bm{x},t))$, such that $|\Psi_0|^2=\rho_0$, $\langle \psi(\bm{x},t) \rangle=0$ and $|\psi|\ll 1$, and defines the Fourier series as $\psi(x)=\Sigma_\bfk \psi_\bfk e^{i\bfk \cdot \bfx}$. We obtain $a_\bfk$ from $\psi_\bfk$ by finding a linear Bogoliubov transformation, and this allows us to find the relationship between the particle spectrum (ignoring the zero mode), defined as
\begin{equation}\label{eq:ntildek}
n_\bfk=\left( \frac{L}{2\pi}\right)^3 \langle |\Psi_\bfk|^2 \rangle/m=\left( \frac{L}{2\pi}\right)^3 \langle |\psi_\bfk|^2 \rangle\rho_0/m \,,
\end{equation}
and the wave-action spectrum $\nB_\bfk$. One should note that we define the particle spectrum so that $N=\int 4\pi k^2 n_k\rmd k$, where $N$ is the mean particle density.
A different convention was used in \cite{dogra2023universal}: the particle spectrum was defined in the way that $N=\frac{1}{(2\pi)^3}\int 4\pi k^2 n_k\rmd k$. This difference in notation has to be taken into account when dealing with the experimental data.

In the acoustic limit, $\psi_\bfk\to \frac{1}{2\xi\sqrt{\cs k}}(a_\bfk-a^*_{-\bfk})$. 
Assuming that $a_\bfk$ and $a_{-\bfk}$ have the same amplitude and independent random phases  (consistent with the so-called random amplitude and phases assumption in WWT plus the isotropy assumption), one gets 
\begin{equation}
 n_\bfk=\frac{\rho_0}{2m\xi^2\cs k}\nB_\bfk\,.   
\end{equation}
In the short-wave limit, 
we have $\psi_\bfk\to  \frac{1}{\sqrt{\sqrt{2}\cs\xi}}a_\bfk$, which results in 
\begin{equation}\label{eq:nkshift}
    n_\bfk=\frac{\rho_0}{\sqrt{2}m\cs\xi}\nB_\bfk\,.
\end{equation} 

To obtain isotropic WKE, we calculate the angular average of the Dirac-$\delta$ function of wavevectors in \eqref{eq:WKE}, $\Delta=\langle \delta(\bfk-\bfk_1-\bfk_2)\rangle$ in SM, yielding $\Delta=\frac{2\pi}{k k_1 k_2}\,$.  
Remarkably, we discover that for the degenerate resonant manifold of non-dispersive sound waves, 
$\Delta_{\rm nondisp}=\frac{\pi}{kk_1k_2}$, 
which regularizes the 3D non-dispersive isotropic WKE of acoustic turbulence. Nonetheless, to make sense of the WKE, one should take the limit of the large box first, followed by the angle average, and finally the large time limit (small nonlinearity). 
The resulting isotropic 3D acoustic WKE has the same form as \eqref{eq:wke_acous}, except that the prefactor is now $2\pi^2V_0^2/\cs$.
In SM, we present a revised, rigorous derivation of the WKE \eqref{eq:wke_disp},  and the parametrization of the resonant manifold defined by \eqref{eq:resonance}. 

We seek for stationary power-law solutions of wave-kinetic equations using the standard process that the WTT offers \cite{nazarenko2011wave}. We provide all the technical information in SM,
including the calculation and locality analysis, while the main text outlines the essential steps. 
Note that for the non-dispersive acoustic case, we have $P(k)=8\pi^3V_0^2A^2\frac{I(x)}{2x-9}k^{9-2x}$, in which the non-dimensional collision integral $I(x)$ is the same as the one for dispersive acoustic waves, and  obtain  
\begin{equation}
    C_{\rm 1,nondisp}= \frac{\sqrt{2}}{\sqrt{3\pi(4\ln{2}+\pi-1)}}\,.
\end{equation}

\begin{table}[!h]
\centering
\resizebox{\columnwidth}{!}{%
\begin{tabular}{|c|c|c|c|c|c|c|c|}
\hline
run& model& $N_r$ & \multicolumn{2}{|c|}{$k_{\text{min}}$} &$k_{\text{max}}$  & $k_f$ & $s$ \\
\hline
1 & Bogoliubov WKE &$32768$ &\multicolumn{2}{|c|}{$1\times 10^{-2}/\xi$} & $200/\xi$  &  0 &  4 \\
\hline
2 & acoustic WKE  &8196  &\multicolumn{2}{|c|}{$1\times 10^{-4}$} & $1$   & 0.01 & 2 \\
\hline
run& model  &\multicolumn{3}{|c|}{$f_0$} & $\Delta k_f$ & $\beta$& $k_d$  \\
\hline
1 & Bogoliubov WKE &\multicolumn{3}{|c|}{$1.032\times10^{14}$} & $1.5\times 10^{-2}$  & 1& 80 \\
\hline
2 & acoustic WKE &\multicolumn{3}{|c|}{$2.419\times10^{16}$} & 0.028  & 1& 0.6 \\
\hline
run&model &$\xi$ &$N_p$ & $\alpha$ & $k_R$  & $f_{\rm w}^2$ & $\gamma$  \\
\hline
3  &  GPE & $1.25\Delta x$ & 960 &  3 & 260  &  $10^{-9}$ & $-1$  \\
\hline
4  &  GPE & $15\Delta x$  & 576 & 4 & 300 & $10^{-4}$ & $-1$    \\
\hline
5  &  GPE & $1.25\Delta x$ & 384 &  3 & 128 & --- & ---  \\
\hline
6  &  GPE  & $10\Delta x$ & 384 & 4  & 200 & --- & ---  \\
\hline
\end{tabular}
}
\caption{\label{tab:param}%
Parameters for GPE and WKE simulations.} 
\vspace{-4ex}
\end{table}

\subsection*{Numerical simulation methods}
We perform GPE simulations using the standard massively-parallel pseudo-spectral code FROST \cite{KrstulovicHDR} with a fourth-order Exponential Runge-Kutta temporal scheme \cite{zhu2022testing}. 
We use grids of $N_p^3$ collocation points so that the grid length is $\Delta x=L/N_p$. 
The forced-dissipated GPE \eqref{eq:GP} is used for the GPE runs 3 and 4, 
with stochastic forcing term supported on a narrow band ($1\le k \le 3$) and obeying the Ornstein–Ulenbeck process  $\mathrm{d}\mathcal{F}_{\bf k} (t)=-\gamma\, \widehat{\Psi}_{\bf{k}}\mathrm{d}t+f_{\rm w}\mathrm{d}{W}_{\bf k}$ where {$W_{\bf k}$} is the Wiener process. The parameters $\gamma$ and $f_0$ control the correlation time and the amplitude of the forcing, respectively.  
In addition, the condensate amplitude (at $k=0$) is kept constant during evolution. 
Dissipation is implemented through the operator $\mathcal D$ in Eq.~\ref{eq:GP}. In Fourier space, its action is
$\widehat{\mathcal D}(\mathbf{k}) = D_k \hat{\Psi}_{\mathbf{k}},$ where $D_k=\left(\frac{k}{k_R}\right)^{2\alpha},$ representing a smooth hyper-viscous damping acting at small scales.
 Finally, the $k$-space energy fluxes $P(k)$ are computed directly using the GPE \eqref{eq:GP} (see Supplemental Materials of \cite{griffin2022energy}). We verify that the simulated WT is in a weakly nonlinear regime by measuring spatio-temporal spectra of $\Psi({\bf x}, t)$ and confirming that they are narrowly concentrated around the dispersion curve \eqref{eq:Bogodis} (see SM). 

To simulate BEC with a potential trap, we set a box trap and release the constraint on the $k=0$ mode. The potential term reads $V(\bm x)= 0$ for coordinates satisfying $|x|, |y|, |z| \le L_{\rm trap}$, and $V_{\rm box}(\bm{x}) = U_D$ otherwise.
$U_D=1000\mu$ is the value we select for the GPE runs 5 and 6. $\mu=mc_s^2$ is the chemical potential.  We obtain the ground state using the imaginary-time scheme with this box trap.
We then add energy by replacing the forcing term $i\mathcal{F}$ in \eqref{eq:GP} by an shaking potential term $V_{osc}(\bfx,t)$, namely, $i\mathcal{F}=V_{osc}(\bfx,t)=A\,x \sin(\omega_{\rm res} t)+ A\, y \cos(\frac{\sqrt{5}}{2}\omega_{\rm res} t)$. Note that we shake the trap in both $x$ and $y$ directions to isotropize the system, which is particularly necessary to obtain acoustic waves at relatively large scales. 
The factor $\sqrt{5}/2$ is chosen so that the ratio of the shaking frequencies in the $x$ and $y$ directions is irrational. This makes the drivings in the two directions effectively independent, while keeping the frequencies as close as possible.
The shaking amplitude is set as $A=0.5 \mu/L_{\rm trap}$ and the frequency is set as the Bogoliubov frequency of the trap, which is $\omega_{\rm res}=c_s k_{\rm trap}\sqrt{1+(k_{\rm trap}\xi)^2/2}$. Dissipation is set in the same way as in the periodic case, but the condensate amplitude is not fixed. 
The energy flux is estimated under the stationary-state assumption from the total dissipation by 
$P_0=2 \int D_k E(k) \rmd k$. This provides a practical flux normalization for the weak-wave-turbulence-dominated regime considered here, especially in the high-$k$ range where the dissipation acts and the nonlinear contribution is expected to remain very weak.
The density, sound speed, and healing length used in Fig.~\ref{fig:EkPk}(a) for runs 5 and 6 are the effective values inside the trap when the system gets stationary.

 We simulate isotropic WKE using the code WavKinS \cite{WavKinS}. Similarly to the GPE, we also include a forcing term of the form $f_k=f_0 k^s e^{-(k-k_{\rm f})^2/\Delta k_{\rm f}^2 }$ and dissipation term $-(k/k_d)^{\beta} n_k$ in the right-hand side of WKE \eqref{eq:wke_disp} and \eqref{eq:wke_acous}. 
The code uses a logarithmic mesh $\{k_i=C\lambda^i, i=1,\ldots,N_r\}$, where $\lambda$ is chosen in order to span the integration domain $[k_{\rm min},k_{\rm max}]$. Inter-mesh values are obtained using a linear interpolation that ensures positivity. 
The collision integral is a 1D integral obtained by making explicit use of the parameterization of the resonant manifold provided in SM. 
Integrals are performed using the trapezoidal rule, after performing a change of variable in the following manner $\int\limits_{k_i}^{k_{i+1}} G(p) \rmd p  \approx \frac{\ln{\lambda}}{2}\left(G(k_{i+1})k_{i+1}+G(k_{i})k_{i}\right)$.
We make an approximation of $\int\limits_{0}^{k_{1}} G(p) \rmd p\approx k_1 G(k_1)$ for the integral on the first bin. Finally, we use a Runge-Kutta-2 time marching method, with an implicit scheme for the dissipative term.

All essential numerical parameters for the WKE and GPE runs are given in
Table~\ref{tab:param}. 
The choice of parameters in each run were made to maximize the inertial (non-dissipative) range of $k$ in the regime at which this specific run is focused: large-scale acoustic, small-scale Bogoliubov, or the regime containing both large- and small-scale WT.

\subsection*{EoS predicted by 4-wave and 3-wave WT and fitting method}
In reference \cite{zhu2023direct}, the constant energy flux solution in the $4$-wave regime was derived analytically and confirmed numerically. Its $\log$-corrected spectrum reads
\begin{equation}
    n_k=C_\rmd \left(\epsilon \hbar/g^2 \right)^{1/3}k^{-3}\ln^{-1/3}(k/k_{\rm f})\,.
\end{equation}
Substituting $g=4\pi \hbar^2 a_s/m$ into the above equation, one gets the equation of states of \eqref{eq:EoS4}.

We recall the relationship between the particle spectrum and the wave-action spectrum for short Bogoliubov waves \eqref{eq:nkshift}, and substitute $\nB_k=\frac{E(k)}{4\pi k^2\omega_{\rm short}(k)}$, the KZ solution \eqref{eq:KZshort}, and $P_0=\epsilon/\rho_0$ into it. This gives
\begin{equation}\label{eq:ntildek3}
    n_k=\frac{C_2}{4\pi}\frac{\rho_0^{1/2}\xi^{1/2}}{m\cs^{3/2}}\epsilon^{1/2}k^{-3}\,.
\end{equation}
Then we express the parameters $\cs$, $\xi$ and $\rho_0$ in terms of $\hbar$, $m$, $a_s$ and $N$ 
by the definitions  $\xi=\sqrt{\frac{\hbar^2}{2g\rho_0}}$, $\cs=\sqrt{\frac{g\rho_0}{m^2}}$, $g=4\pi \hbar^2 a_s/m$, and $\rho_0=mN$. Finally, one gets 
\begin{equation}
n_k=\frac{C_2}{2^{17/4}\pi^2} \frac{m \epsilon^{1/2}}{\hbar^{3/2}N^{1/2}a_s}k^{-3}\,,
\end{equation}
and consequently the EoS \eqref{eq:EoS3}.

\begin{figure}
\centering
  \includegraphics[width=0.48\textwidth]{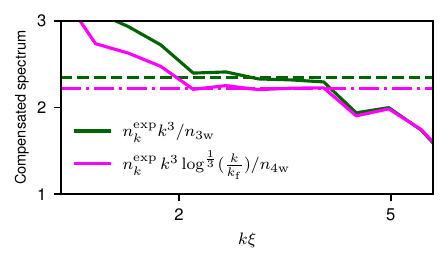}
  \caption{\label{fig:Four-Three-wave} 
Experimental spectrum $n^{\rm exp}_k$ compensated by $k^3 \ln^{1/3}(\frac{k}{k_{\rm f}})$ and $k^3$, respectively, corresponding to the leftmost point in Fig.~\ref{fig:EoS} (a). This provides an example in which the 4-wave prediction is selected.
\vspace{-2ex}
}
\end{figure}

In generating Fig.~\ref{fig:EoS}, each experimental spectrum is fitted using the 4-wave prediction $n_k = n_{\rm 4w}^{\rm exp} k^{-3}\ln^{-\frac{1}{3}}\left(\frac{k}{k_{\rm f}}\right)$ and the 3-wave predictions $n_k = n_{\rm 3w}^{\rm exp} k^{-3}$, respectively, yielding the experimentally obtained spectrum amplitude $n^{\rm exp}_{\rm 4w}$ and $n^{\rm exp}_{\rm 3w}$. To determine the most suitable candidate between these two models, theoretical predictions are calculated $n_{\rm 4w}$ and $n_{\rm 3w}$ using \eqref{eq:EoS4} and \eqref{eq:EoS3}. The model is then selected based on the computed ratio $n^{\rm exp}_{\rm 4w}/n_{\rm 4w}$ and $n^{\rm exp}_{\rm 3w}/n_{\rm 3w}$, prioritizing the one whose ratio is closes to unity. 
An explicit example in which the 4-wave prediction is selected is shown in Fig.~\ref{fig:Four-Three-wave}, corresponding to the left-most data point in Fig.~\ref{fig:EoS}(a).
Note that the vertical-axis scale in Fig.~\ref{fig:EoS}(a) differs from that in Fig.~4(a) of \cite{dogra2023universal}, because a different convention for \(n(k)\) is used here, differing by a factor of \(1/(8\pi^3)\), as explained earlier in the Methods.

\begin{acknowledgments}
This work was funded by the Simons Foundation Collaboration grant Wave Turbulence (Award ID 651471). 
This work was provided with computing HPC and storage resources of IDRIS and CINES under the allocations A0152A14637 and A0172A14637 made by GENCI on the
Jean Zay (SKL partition) and Adastra (GENOA partition) supercomputers. 
This work was granted access to the
OPAL infrastructure from Université Côte d’Azur, supported
by the French government, through the UCAJEDI Investments in the Future project managed by the National Research
Agency (ANR) under Reference No. ANR-15-IDEX-01.
Y. Z. and S. N. acknowledge the support from ANR through the 
project VORTECS (Grant No. ANR-22-CE30-0011).
\end{acknowledgments}

\bibliographystyle{apsrev4-1}
\bibliography{WWTT}

@article{moreno2025observation,
  title={Observation of Relaxation Stages in a Nonequilibrium Closed Quantum System: Decaying Turbulence in a Trapped Superfluid},
  author={Moreno-Armijos, MA and Fritsch, Amilson Rogelso and Garc{\'\i}a-Orozco, AD and Sab, Sarah and Telles, G and Zhu, Ying and Madeira, Lucas and Nazarenko, Sergey and Yukalov, Vyacheslav I and Bagnato, Vanderlei Salvador},
  journal={Physical Review Letters},
  volume={134},
  number={2},
  pages={023401},
  year={2025},
  publisher={APS}
}

@misc{WavKinS,
  title        = {WavKinS: Wave Kinetic Solver },
  author       = {Krstulovic, Giorgio and Labarre, Vincent},
  note= {Currently under development.},
   url = {https://gkrstulovic.gitlab.io/project/wavkins/},
  year         = 2024
}

@article{kochurin2024three,
  title={Three-Dimensional Acoustic Turbulence: Weak Versus Strong},
  author={Kochurin, EA and Kuznetsov, EA},
  journal={arXiv preprint arXiv:2407.08352},
  year={2024}
}

@misc{martirosyan2024_EoS_GP,
      title={An Equation of State for Turbulence in the Gross-Pitaevskii model}, 
      author={Gevorg Martirosyan and Kazuya Fujimoto and Nir Navon},
      year={2024},
      eprint={2407.08738},
      archivePrefix={arXiv},
      primaryClass={cond-mat.quant-gas},
      url={https://arxiv.org/abs/2407.08738}, 
}

@article{Polanco_VortexClusteringPolarisation_2021,
 author = {Polanco, Juan Ignacio and Müller, Nicolás P. and Krstulovic, Giorgio},
 date = {2021-12},
 doi = {10.1038/s41467-021-27382-6},
 issn = {2041-1723},
 journal = {Nature Communications},
 number = {1},
 year = {2021},
 pages = {7090},
 shortjournal = {Nat Commun},
 title = {Vortex Clustering, Polarisation and Circulation Intermittency in Classical and Quantum Turbulence},
 url = {https://www.nature.com/articles/s41467-021-27382-6},
 volume = {12}
}

@article{Zakharov_CollapseLangmuirWaves_,
  title = {Collapse of {{Langmuir Waves}}},
  author = {Zakharov, V E},
  year = {1972},
  month = aug,
  journal = {Soviet Physics-JETP},
  volume = {35},
  pages = {908--914},
  pages = {7},
  langid = {english}
}

@article{galtier2017turbulence,
  title={Turbulence of weak gravitational waves in the early universe},
  author={Galtier, S{\'e}bastien and Nazarenko, Sergey V},
  journal={Physical review letters},
  volume={119},
  number={22},
  pages={221101},
  year={2017},
  publisher={APS}
}

@article{Caillol_KineticEquationsStationary_2000,
  title = {Kinetic Equations and Stationary Energy Spectra of Weakly Nonlinear Internal Gravity Waves},
  author = {Caillol, P. and Zeitlin, V.},
  year = {2000},
  month = jul,
  journal = {Dynamics of Atmospheres and Oceans},
  volume = {32},
  number = {2},
  pages = {81--112},
  issn = {03770265},
  doi = {10.1016/S0377-0265(99)00043-3},
  langid = {english}
}

@article{Lvov_WeakTurbulenceKelvin_2010,
  title = {Weak Turbulence of {{Kelvin}} Waves in Superfluid {{He}}},
  author = {L'vov, Victor S. and Nazarenko, Sergey},
  year = {2010},
  month = aug,
  journal = {Low Temperature Physics},
  volume = {36},
  number = {8},
  pages = {785--791},
  issn = {1063-777X, 1090-6517},
  doi = {10.1063/1.3499242},
  langid = {english}
}

@article{Galtier_WeakInertialwaveTurbulence_2003,
  title = {Weak Inertial-Wave Turbulence Theory},
  author = {Galtier, S{\'e}bastien},
  year = {2003},
  month = jul,
  journal = {Physical Review E},
  volume = {68},
  number = {1},
  pages = {015301},
  issn = {1063-651X, 1095-3787},
  doi = {10.1103/PhysRevE.68.015301},
  langid = {english}
}

@book{landau1987fluid,
  title={Fluid Mechanics: Volume 6},
  author={Landau, LD and Lifshitz, EM},
  volume={6},
  year={1987},
  publisher={Elsevier}
}

@ARTICLE{during2009,
       author = {{D{\"u}ring}, Gustavo and {Picozzi}, Antonio and {Rica}, Sergio},
        title = "{Breakdown of weak-turbulence and nonlinear wave condensation}",
      journal = {Physica D Nonlinear Phenomena},
         year = 2009,
        month = aug,
       volume = {238},
       number = {16},
        pages = {1524-1549},
          doi = {10.1016/j.physd.2009.04.014},
       adsurl = {https://ui.adsabs.harvard.edu/abs/2009PhyD..238.1524D}
}

@article{dogra2023universal,
  title={Universal equation of state for wave turbulence in a quantum gas},
  author={Dogra, Lena H and Martirosyan, Gevorg and Hilker, Timon A and Glidden, Jake AP and Etrych, Ji{\v{r}}{\'\i} and Cao, Alec and Eigen, Christoph and Smith, Robert P and Hadzibabic, Zoran},
  journal={Nature},
  volume={620},
  number={7974},
  pages={521--524},
  year={2023},
  publisher={Nature Publishing Group UK London}
}

@article{zhu2022testing,
  title={Testing wave turbulence theory for the Gross-Pitaevskii system},
  author={Zhu, Ying and Semisalov, Boris and Krstulovic, Giorgio and Nazarenko, Sergey},
  journal={Physical Review E},
  volume={106},
  number={1},
  pages={014205},
  year={2022},
  publisher={APS}
}

@article{zhu2023direct,
  title={Direct and inverse cascades in turbulent Bose-Einstein condensates},
  author={Zhu, Ying and Semisalov, Boris and Krstulovic, Giorgio and Nazarenko, Sergey},
  journal={Physical Review Letters},
  volume={130},
  number={13},
  pages={133001},
  year={2023},
  publisher={APS}
}

@article{zakharov1970spectrum,
  title={Spectrum of acoustic turbulence},
  author={Zakharov, Vladimir Evgen'evich and Sagdeev, Roal'd Zinnurovich},
  journal={Doklady Akademii Nauk},
  volume={192},
  number={2},
  pages={297--300},
  year={1970},
  publisher={Russian Academy of Sciences}
}

@article{griffin2022energy,
  title={Energy spectrum of two-dimensional acoustic turbulence},
  author={Griffin, Adam and Krstulovic, Giorgio and L’vov, Victor S and Nazarenko, Sergey},
  journal={Physical Review Letters},
  volume={128},
  number={22},
  pages={224501},
  year={2022},
  publisher={APS}
}

@article{Dispersive,
  title = {Dispersive Bottleneck Delaying Thermalization of Turbulent Bose-Einstein Condensates},
  author = {Krstulovic, Giorgio and Brachet, Marc},
  journal = {Phys. Rev. Lett.},
  volume = {106},
  issue = {11},
  pages = {115303},
  numpages = {4},
  year = {2011},
  month = {Mar},
  publisher = {American Physical Society},
  doi = {10.1103/PhysRevLett.106.115303},
  url = {https://link.aps.org/doi/10.1103/PhysRevLett.106.115303}
}

@article{navon2016emergence,
  title={Emergence of a turbulent cascade in a quantum gas},
  author={Navon, Nir and Gaunt, Alexander L and Smith, Robert P and Hadzibabic, Zoran},
  journal={Nature},
  volume={539},
  number={7627},
  pages={72--75},
  year={2016},
  publisher={Nature Publishing Group}
}

@book{nazarenko2011wave,
  title={Wave turbulence},
  author={Nazarenko, Sergey},
  volume={825},
  year={2011},
  publisher={Springer Science \& Business Media}
}

@article{ZLF,
  title="{Kolmogorov spectra of turbulence 1: Wave turbulence}",
  author={Vladimir Evgen'evich Zakharov and Viktor Sergeevich L'vov and Grigory Falkovich},
  year={1992},
  journal={Springer Series in Nonlinear Dynamics},
  publisher={Springer-Verlag}
}

@article{dyachenko1992optical,
  title="{Optical turbulence: weak turbulence, condensates and collapsing filaments in the nonlinear Schr{\"o}dinger equation}",
  author={Dyachenko, S and Newell, AC and Pushkarev, A and Zakharov, VE},
  journal={Physica D: Nonlinear Phenomena},
  volume={57},
  number={1-2},
  pages={96--160},
  year={1992},
  publisher={Elsevier}
}

@phdthesis{KrstulovicHDR,
author={Giorgio Krstulovic},
title="{A theoretical description of vortex dynamics in superfluids. Kelvin waves, reconnections and particle-vortex interaction}",
school={Universite C\^ote d’Azur},
type = {Habilitation \`a diriger des recherches},
month={October},
URL={https://gkrstulovic.gitlab.io/thesisms/hdr-krstulovic/},
year={2020}
}

\clearpage
\appendix
\begin{widetext}
  \begin{center}
    \Large
    \textbf{Turbulence and far-from-equilibrium equation of state of Bogoliubov waves in Bose-Einstein Condensates: Supplementary Materials}
  \end{center}
  \renewcommand{\figurename}{Supplementary Fig.}
  \newcommand\figref[1]{Supplementary Fig.~\ref{#1}}
  \renewcommand\thetable{S\arabic{table}}
  \setcounter{figure}{0}
  \setcounter{table}{0}

  \setcounter{equation}{0}
  \renewcommand\theequation{S\arabic{equation}}

\section{Derivation of wave-kinetic equations}  \label{append:a}

In this Supporting Information, we first review the Hamiltonian formulation of Bogoliubov waves following the methods in \cite{griffin2022energy, dyachenko1992optical} to obtain the interaction term represented by  $V^{1}_{ {23}}$ in Eq.~(8) of the main text.  We also validate $V^{1}_{ {23}}$ by using the Bogoliubov transformation. The angular average of the Dirac-$\delta$ function in terms of wavevectors is discussed, which is essential to getting the isotropic WKE.
Specifically, we clarify that even in non-dispersive 3D acoustic systems, the singularity of WKE is canceled by taking an angular average of $\delta(\bfk-\bfk_1-\bfk_2)$, which results in a prefactor of the corresponding WKE that is different from the one obtained for the dispersive case in the limit of vanishing dispersion.

\subsection{Hamiltonian formation and wave interaction amplitude $V^1_{23}$}

The action for Bogoliubov waves (per unit of mass) is expressed in hydrodynamic variables it is given by
\begin{align}\label{eq:action}
  \mathcal{S} = \frac{1}{\rho_0 L^3}\int dt d^3x \left[ -\rho \dot{\phi}  - \frac{\rho}{2} (\nabla \phi)^2 - \frac{c_s^2}{2\rho_0}(\rho - \rho_0)^2 -\cs^2\xi^2\left( \nabla\sqrt{\rho}   \right)^2  \right], 
\end{align}
which corresponds to a compressible, isentropic, irrotational fluid \cite{landau1987fluid} but with an extra quantum pressure term $\cs^2\xi^2\left( \nabla\sqrt{\rho}   \right)^2$. $L$ is the length of the 
triply-periodic box on which the problem is defined. 

Following the change of variables defined in the SM of \cite{griffin2022energy}, i.e., $\rho=\rho_0(1+A)^2$ and $p=2(1+A)\phi$, and applying a similar procedure, we get the second- and the third-order terms of the Hamiltonian,
\begin{equation}
   H_2  = \int \frac{d^3 x}{L^3} \left[ \frac{1}{8} (\nabla p)^2 + 2\cs^2 A^2+\cs^2\xi^2(\nabla A)^2 \right] ,\quad
   H_3  = \int \frac{d^3x}{L^3} \left[2\cs^2A^3 - \frac{1}{4} p(\nabla p)\cdot (\nabla A) \right]\,,
\end{equation}
with the total Hamiltonian being $H\approx H_2+H_3$.
By defining the Fourier series as $p({\bf x})=\sum_{\bf k} p_{\bf k}e^{i {\bf k}\cdot{\bf x}}$ and $A({\bf x})=\sum_{\bf k} A_{\bf k}e^{i {\bf k}\cdot{\bf x}}$, $H_2$ and $H_3$ become:
\begin{equation}
  H_2  = \sum_\bfk \frac{1}{8}k^2 |p_\bfk|^2 + 2\omega_k^2k^{-2}|A_\bfk|^2, \quad
  H_3  = \sum_{1,2,3} 2 \cs^2A_1 A_2 A_3 \delta_{1,2,3} + \frac{1}{4}p_1 p_2 A_3 \bfk_2 \cdot \bfk_3 \delta_{1,2,3},
\end{equation} 
 where $\omega_k=\cs k \sqrt{1+(k\xi)^2/2}$ is the Bogoliubov dispersion relation, and $\delta_{1,2,3}$ is $1$ if $\bfk_1+\bfk_2+\bfk_3=0$, and $0$ otherwise.

 To write the Hamiltonian in the canonical form, we perform the following change of variables,
\begin{equation}
    p_\bfk=i\sqrt{2}\omega_k^{1/2}k^{-1}(a_\bfk-a^*_{-\bfk})\,,\quad  A_\bfk=(2\sqrt{2})^{-1}\omega_k^{-1/2}k(a_\bfk+a^*_{-\bfk})\,,
\end{equation} 
so that $H_2=\sum_\bfk \omega_k a_\bfk a^*_\bfk$ and $i \dot{a_\bfk}=\frac{\partial H}{\partial a^*_\bfk}$ in the leading order. Using the resonance condition that $\bfk_1+\bfk_2+\bfk_3=0$, $H_3$ reduces to 
\begin{equation} \label{eq:H3}
    H_3=\sum_{1,2,3} \frac{1}{8\sqrt{2}}\cs^{-1}\left( \omega_1\omega_2\omega_3 \right)^{1/2}   \left[ 
    \frac{3}{\eta_1\eta_2\eta_3}+(k_1k_2k_3)^{-1}\left( \frac{k_1^3}{\eta_1} - \frac{k_2^3}{\eta_2}- \frac{k_3^3}{\eta_3}
 \right)\right]\left( a_1a^*_2a^*_3 +\text{c.c} \right) \delta^1_{2,3}=\sum_{1,2,3}V^1_{2,3}\left( a_1a^*_2a^*_3 +\text{c.c} \right)\delta^1_{2,3}\,,
\end{equation}
 where $\eta_i\equiv\eta(k_i) = \sqrt{1+(k_i\xi)^2/2}$, $\omega_i \equiv\omega(k_i) $ for $i=1,2,3$, and $\delta^1_{2,3}=\delta_{-1,2,3}$.

 Equation \eqref{eq:H3} gives the expression of $V^1_{23}$ as in Eq.~(8) of the main text. Note that by using the following identity provided by the frequency resonance, 
 \begin{equation}
     \frac{k_1^3}{\eta_1} - \frac{k_2^3}{\eta_2}- \frac{k_3^3}{\eta_3}
     =2\left( \frac{k_2}{\eta_2\xi^2} +  \frac{k_3}{\eta_3\xi^2}- \frac{k_1}{\eta_1\xi^2} \right)\,,
 \end{equation}
we get an equivalent expression for $V^1_{23}=V_0\sqrt{k_1k_2k_3}W^1_{23}$ with 
\begin{equation}\label{eq:W123}
W^1_{23}=\frac{1}{2\sqrt{\eta_1\eta_2\eta_3}} + \frac{\sqrt{\eta_1\eta_2\eta_3}}{3\xi^2k_1k_2k_3}\left(
\frac{k_2}{\eta_2} + \frac{k_3}{\eta_3}- \frac{k_1}{\eta_1}\right) \,.
\end{equation}
The above expression is more convenient  for the case  $k\xi\to \infty$ whereas Eq.~(8) of the main text is better for the $k\xi\to 0$ limit.

\subsection{A different method to obtain the interaction coefficient: the Bogoliubov transformation}
A different way to find the interaction coefficient $V^1_{23}$ exploits the Bogoliubov transformation.
The starting point is the Hamiltonian (per unit of mass) for the  GPE in terms of $\Psi$ and $\rho$. It reads
\begin{equation}
    H=\frac{1}{L^3\rho_0}\int \left(\cs^2\xi^2 |\nabla\Psi|^2+   \frac{\cs^2}{2\rho_0}(\rho-\rho_0)^2\right)\rmd^3x\,.
\end{equation}
Consider weak  disturbances on the background of a strong coherent condensate,  $\Psi(\bm{x},t)=\Psi_0(1+\psi(\bm{x},t))$, such that $|\Psi_0|^2=\rho_0$, $\langle \psi(\bm{x},t) \rangle=0$ and $|\psi|\ll 1$; then the second and third order of $H$ become
\begin{equation}
    H_2=\frac{1}{L^3}\int \left( \cs^2\xi^2|\nabla\psi|^2+\frac{\cs^2}{2}(\psi+\psi^*)^2   \right)\rmd^3x \,,\quad 
    H_3=\frac{1}{L^3}\int \cs^2|\psi|^2(\psi+\psi^*)\rmd^3x \,.
\end{equation}
Write $\psi(\bm{x})=\sum_\bfk \psi_{\bfk} e^{i\bfk\cdot\bm{x}}$;  $H_2$ and $H_3$ become
\begin{equation}
    H_2=\sum_\bfk \cs^2 (1+k^2\xi^2)|\psi_\bfk|^2+\frac{\cs^2}{2}(\psi_\bfk\psi_{-\bfk}+\psi^*_\bfk\psi^*_{-\bfk})\,,\quad 
    H_3=\sum_{1,2,3} \cs^2 \psi_{\bfk_1}\psi^*_{\bfk_2}\psi^*_{\bfk_3}\delta^1_{23}+\text{c.c}\,.
\end{equation}
To kill the non-diagonal terms in $H_2$, we perform the method described in \cite{ZLF} and find a canonical transformation as follows:
\begin{equation} \label{eq:canonical}
    \psi_\bfk=\frac{1}{2}\sqrt{\frac{k}{2\cs \eta_k}}\left[\left( 1+\frac{\sqrt{2}\eta_k}{k\xi} \right) b_\bfk +\left( 1-\frac{\sqrt{2}\eta_k}{k\xi} \right) b^*_{-\bfk}\right]\,,
\end{equation}
and the inverse transformation 
\begin{equation} 
    b_\bfk=\frac{1}{2}\sqrt{\frac{\cs k\xi^2}{\eta_k}}\left[\left( 1+\frac{\sqrt{2}\eta_k}{k\xi} \right) \psi_\bfk +\left( 1-\frac{\sqrt{2}\eta_k}{k\xi} \right) \psi^*_{-\bfk}\right]\,,
\end{equation}
so that $H_2=\sum_\bfk \omega_k b_\bfk b_\bfk^*$, which means $b_\bfk=a_\bfk$.

\subsection{Angular average of $\delta(\bfk-\bfk_1-\bfk_2)$}
 
To get the isotropic WKE, we only need to compute the angular average of the Dirac-$\delta$ function of wavevectors in Eq.~(7) of the main text, since $V^1_{23}$ and frequency are both angle-independent.  
Writing $\rmd \bfk=k^2 \rmd \Omega \rmd k$, where $\rmd\Omega=\sin\theta\rmd\theta\rmd\phi$, the angular average of $\delta(\bfk-\bfk_1-\bfk_2)$ is then defined as $\Delta=\int \delta(\bfk-\bfk_1-\bfk_2)\rmd \Omega_1 \rmd \Omega_2$. 

We define a coordinate system and the spherical angles as shown in \figref{fig:delta}(a).
\begin{figure}[h]
\centering
\includegraphics[width=0.6\textwidth]{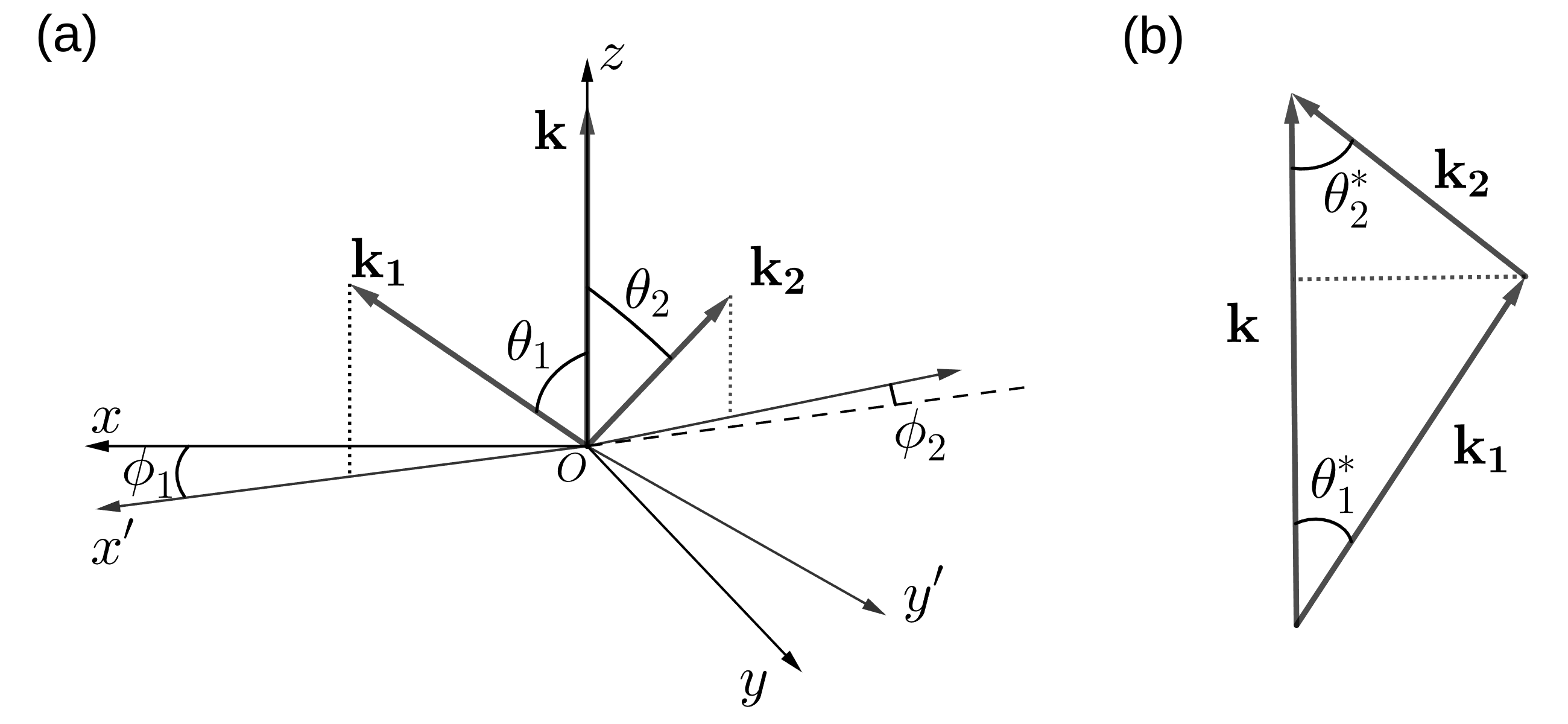}%
\caption{\label{fig:delta}
(a) Coordinates system to compute $\Delta$; (b) Wave vector triad. 
}
\end{figure}
Decomposing $\delta (\bfk-\bfk_1-\bfk_2)$ in these coordinates,
we get
\begin{equation}
\begin{split}
&\Delta=\int\delta(k_1\sin\theta_1+k_2\sin\theta_2\cos(\phi_2+\pi))\delta(k_2\sin\theta_2\sin(\phi_2+\pi))  \delta(k-k_1\cos\theta_1-k_2\cos\theta_2)\sin\theta_1\sin\theta_2\rmd \theta_1 \rmd\theta_2\rmd \phi_1\rmd\phi_2 \\
&=2\pi\int\delta(k_1\sin\theta_1-k_2\sin\theta_2\cos\phi_2)\delta(k_2\sin\theta_2\sin\phi_2) \delta(k-k_1\cos\theta_1-k_2\cos\theta_2)\sin\theta_1\sin\theta_2\rmd \theta_1 \rmd\theta_2\rmd\phi_2. 
\end{split}
\end{equation}
It is easy to find that the system of equations $h_1(\theta_1,\theta_2,\phi_2)\equiv k_1\sin\theta_1-k_2\sin\theta_2\cos\phi_2=0$, $h_2(\theta_1,\theta_2,\phi_2)\equiv k_2\sin\theta_2\sin\phi_2=0$, and $h_3(\theta_1,\theta_2,\phi_2)\equiv k-k_1\cos\theta_1-k_2\cos\theta_2=0$
has a unique solution $\theta_1=\theta_1^*$, $\theta_2=\theta_2^*$ and $\phi_2=0$ for given $k>0$, $k_1\,,k_2\ge0$, in the range $0\le \theta_1\,,\theta_2\le \pi$, and $0\le\phi_2\le 2\pi$.
Consequently, we obtain 
\begin{equation}
\label{Del}
\Delta=\frac{2\pi \sin\theta_1^* \sin\theta_2^*}{|\nabla (h_1, h_2, h_3)| _{(\theta_1^*,\theta_2^*,0)}|}
=2\pi \sin\theta_1^* \sin\theta_2^* /|k_1k_2^2\sin\theta_2^*(\cos\theta_2^*\sin\theta_1^*+\cos\theta_1^*\sin\theta_2^*)|.
\end{equation}
Note that in the resonant triangle as shown in \figref{fig:delta}(b), we have $k_2(\cos\theta_2^*+\cos\theta_1^*\sin\theta_2^*/\sin\theta_1^*)=k$ and we finally  get
\begin{equation}\label{eq:Delta}
    \Delta =\frac{2\pi}{k k_1 k_2}\,.
\end{equation}
Note that for expression \eqref{Del} to be determined,
the vectors $\bfk,\bfk_1$ and $\bfk_2$ cannot be strictly co-linear,  as is the case for non-dispersive sound.
Therefore, in this derivation, it is essential that the system is dispersive. Otherwise, $\Delta$ is $0/0$ undetermined, i.e. the above method originated from \cite{zakharov1970spectrum} fails, as was already stressed in \cite{zakharov1970spectrum}.

Another way to compute $\Delta$ is representing the Dirac-$\delta$ function as
\begin{equation}\label{eq:gauss-delta}
    \delta(\bfk-\bfk_1-\bfk_2)=\frac{1}{(2\pi)^3}\int_{\mathcal{R}^3}\exp[-i\bm{r}\cdot(\bfk-\bfk_1-\bfk_2)]\rmd\bm{r}\,.
\end{equation}
Now we rewrite $\Delta=\frac{1}{4\pi}\int \delta(\bfk-\bfk_1-\bfk_2)\rmd\Omega\rmd\Omega_1\rmd\Omega_2$,
and substitute  \eqref{eq:gauss-delta} into it. Following the method in \cite{zhu2022testing} and using Mathematica, we obtain an analytical expression 
\begin{equation}
    \Delta=\pi\left(\operatorname{sign}(k_1+k_2-k)+ \operatorname{sign}(k+k_1-k_2)+ \operatorname{sign}(k+k_2-k_1)-1\right)/(kk_1k_2). \label{Eq:Deltagen}
\end{equation}
Consider a resonant condition such that $\bfk$, $\bfk_1$, and $\bfk_2$ form a non-degenerated triangle, then, one gets the same $\Delta$ as \eqref{eq:Delta}. 

For the degenerate resonant manifold of non-dispersive sound waves, we have $\bfk||\bfk_1||\bfk_2$, 
$\operatorname{sign}(k_1+k_2-k)=0$, so we get
\begin{equation}
    \Delta_{\rm nondisp}=\frac{\pi}{kk_1k_2}\,.
\end{equation}
Note that the previous term differs in a factor $2$ with respect to Eq.~\ref{eq:Delta} as a consequence of the discontinuous character of formula \eqref{Eq:Deltagen}. 

%

\subsection{Parametrization of the resonant manifold of Bogoliubov waves}

After angular average, the resonant manifold of the wave kinetic equation (9) of the main text takes a much simpler and compact form. Noting that the second and the third terms of the right hand side are the same after permutation of indexes, on is left with only two resonant manifolds determined by the following equations
\begin{equation}
    h_1(k,k_1,k_2)\equiv\omega_1+\omega_2-\omega_k=0,\qquad   h_2(k,k_1,k_2)\equiv-\omega_1 + \omega_2+\omega_k=0,
\end{equation}
for the first and the second integral respectively. One can easily express $k_2$ as a function of $k$ and $k_1$, which for both equations reads
\begin{equation}
    k_2^*=k_2(k,k_1)=\frac{1}{\xi}\left[\sqrt{1+\frac{2 \xi^2}{c_s^2}(\omega_k-\omega_1)^2}- 1\right]^{1/2}.\label{eq:res_manifolds}
\end{equation}
The above solution is valid for $k_1\le k$ in the case $h_1(k,k_1,k_2)=0$, and for $k\le k_1$ in the case $h_2(k,k_1,k_2)=0$. 

Finally, note that to evaluate the integrals of the WKE, one needs to compute
\begin{eqnarray}
    g(k_2)=\left|\frac{\partial{h_1}}{\partial k_2} \right|=  \left|\frac{\partial{h_2}}{\partial k_2} \right|=c_s\frac{1+\xi^2 k_2^2}{\sqrt{1+\frac{1}{2}\xi^2 k_2^2}},
\end{eqnarray}
which is always positive, and therefore, no singularities are present in the Bogoliubov WKE.

Using the previous two equations, we can express the Bogoliubov WKE in a explicit form
\begin{align}\label{eq:reduced_wke}
    &\frac{\partial \nB}{\partial t} 
    = \frac{4 \pi^2 V_0^2}{c_s} \left[ \int_0^k |W^k_{1{2^*}}|^2 \frac{k_1^2 {k_2^*}^2}{g(k_2^*)}\mathcal{T}^k_{1{2^*}}  \rmd k_1  
    -2 \int_k^\infty |W^1_{k2^*}|^2 \frac{k_1^2 {k_2^*}^2}{g(k_2^*)}\mathcal{T}^1_{k2^*}  \rmd k_1 \right],
\end{align}
where we recall that $\mathcal{T}^k_{12} = \nB_{k_1} \nB_{k_2} - \nB_k \nB_{k_1} - \nB_k \nB_{k_2} $, and that $k_2^*$ is given by Eq.~\eqref{eq:res_manifolds}.

\section{Derivation of Kolmogorov-type spectra}  \label{append:b}

In this section, we derive the stationary Kolmogorov-type spectra for the acoustic and short-wave limits, including convergence analysis of the collision integral and calculation of constants. 

\subsection{Zakharov-Sagdeev spectrum}
Seeking a power-law solution in the form of $\nB_k=Ak^{-x}$, and substituting it into Eq.~(10) of the main text, the right hand side  becomes $\St_k=4\pi^2V_0^2\cs^{-1}A^2k^{5-2x}I(x)$ with the dimensionless collision integral 
\begin{eqnarray}
\label{eq:Iacs}
 \nonumber   I(x)&=&\int\limits_{q_1\,,q_2\ge 0} (q_1q_2)^{2-x} \left( (1-q_1^x-q_2^x)\delta(1-q_1-q_2) -(q_1^x-q_2^x-1)\delta(q_1-q_2-1) -(q_2^x-q_1^x-1)\delta(q_2-q_1-1)  \right)      \rmd q_1 \rmd q_2\,\\
    &=& \int_0^1 (q (1-q))^{2-x}  (1-q^x-(1-q)^x)\rmd q -2 \int_1^\infty (q (q-1))^{2-x} (q^x-(q-1)^x-1)\rmd q.   
\end{eqnarray}
Note that the second integral is convergent at $q=\infty$ for $x>4$. However, for $x>3$, the first integral diverges at $q=0$ and $q=1^-$ (the integral is symmetric with respect to $q=1/2$) and the second integral diverges too at $q=1^+$. However, the divergences at $q=0$ and $q=1^\pm$ cancel out, which extends the region of convergence. The expression for the collisional integral thus needs to be understood as a principal value integral. To give a convergent expression, we exploit the symmetry of the first integral and shift the second one to rewrite the collision integral as 
\begin{equation}
    I(x)=2\left(\int_0^{1/2}\left(f_1(q,x)-f_2(q,x)\right)\rmd q - \int_{1/2}^{\infty}f_2(q,x)\rmd q   \right)\,,
\end{equation}
where $f_1(q,x)=\left(q(1-q)\right)^{2-x}\left(1-q^x-(1-q)^x\right)$ and $f_2(q,x)=\left(q(1+q)\right)^{2-x}\left((1+q)^x-q^x-1\right)$. Note that $f_1(q,x)-f_2(q,x) \propto q^{4-x}$ when $q\to 0$, so the region of convergence is extended to $x<5$. As before, $f_2(q,x)\propto q^{3-x}$ when $q\to \infty$, which gives us the convergence interval $4<x<5$. $I(x)$ is also convergent at $x=1$ with $I(1)=0$, which corresponds to the thermal equilibrium solution.

Using Mathematica, within the window of convergence, we obtain explicitly 
\begin{equation}\label{eq:IacA}
I(x)=2^{2 x-5} \,
   _2\tilde{F}_1(x-2,2 x-5;2
   (x-2);-2) \Gamma (2
   x-5)+\frac{4}{(x-5) (x-4)
   (x-3)}+\frac{1}{2}
   B(3-x,3-x)-(-1)^x
   B_{-\frac{1}{2}}(3-x,3-x)\,,
\end{equation}
in which $_2\tilde{F}_1(a,b;c;z)=_2F_1(a,b;c;z)/\Gamma(c)$ is the regularized hypergeometric function with $_qF_p(a,b;c;z)$ the hypergeometric function and $\Gamma(c)$ the Euler Gamma function, $B(a, b)$ is the Euler beta function, and $B_z(a,b)$ is the incomplete beta function.
In \figref{fig:Ix}(a) we plot $I(x)$ in its convergence range. 
\begin{figure*}[h]
\centering
\begin{subfigure}[t]{0.46\textwidth}
\centering
\includegraphics[width=0.6\textwidth]{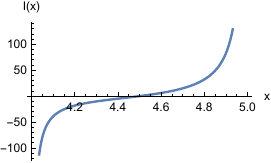}%
\caption{ }
\end{subfigure}
\begin{subfigure}[t]{0.46\textwidth}
\centering
\includegraphics[width=0.6\textwidth]{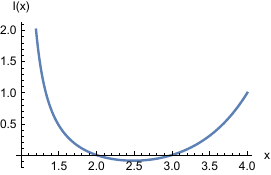}%
\caption{ }
\end{subfigure}
\caption{Dimensionless collision integrals $I(x)$ in their convergence windows.  \textbf{(a)} Collisional integral in the $3$-wave acoustic regime given by Eq.~\eqref{eq:IacA}. We note that $I(9/2)=0$ and $I'(9/2)$ is finite and positive proving that $x^{\rm KZ}=9/2$ solution is local. \textbf{(b)} Collisional integral in the $3$-wave short-wave Bogoliubov regime given by Eq.~\eqref{eq:Ishort}. Note that both, $x^{\rm RJ}=2$ and the $x^{\rm KZ}=3$, are local solutions.\label{fig:Ix}}
\end{figure*}

Now that we have computed the dimensionless collisional integral, we can provide the full expression for the energy spectrum. By definition, $P(k)=-\int_0^k 4\pi\,\tilde{k}^2\omega(\tilde{k})\St(\tilde{k})\rmd \tilde{k}$ so that for an homogeneous waveaction $\nB_k=Ak^{-x}$ it becomes $P(k)=16\pi^3V_0^2A^2\frac{I(x)}{2x-9}k^{9-2x}$. The stationarity of the spectrum means constant flux $P=P_0$ --- independent of $k$, which suggests that $x=9/2$ and $I(9/2)=0$. In such case, one gets $P_0=8\pi^3V_0^2A^2I'(9/2)$ by using the L'H\^{o}pital rule, where $I'(9/2)=\rmd I(x)/\rmd x|_{x=9/2}$. Consequently, we have the stationary spectrum $\nB_k=\sqrt{P_0/(8\pi^3V_0^2I'(9/2))} k^{-9/2}$ and $E(k)=4\pi\cs\sqrt{P_0/(8\pi^3V_0^2I'(9/2))}k^{-3/2}$. We note that $4<9/2<5$ so the exponent of the KZ solution is within the convergence interval of $I(x)$. Furthermore, $I(9/2)=0$ and $0<I'(9/2)<\infty$ (see \figref{fig:Ix}) and therefore this KZ solution is local and mathematically valid. The value of $I'(9/2)$ can be directly computed and reads $I'(9/2)=64(4\ln{2}+\pi-1)/3$. Finally, using this value, we obtain the Zakharov-Sergeev spectrum with the constant as in Eq.~(2) of the main text.


\subsection{Kolmogov-Zakharov spectrum for short Bogoliubov waves}

For short waves, similarly to the procedure performed on acoustic waves, we study the convergence of the integral of Eq.~(11) of the main text and compute the derivative $I'(3)$.  We rewrite the collision integral as
\begin{equation}
    I(x)=\int_0^1g_1(q,x)\rmd q -2 \int_{0}^{\infty}g_2(q,x)\rmd q  \,,
\end{equation}
where $g_1(q,x)=\frac{1}{2}q^{1-x}(1-q^2)^{-x/2}\left(1-q^x-(1-q^2)^{-x/2}\right)$, and  $g_2(q,x)=\frac{1}{2}q^{1-x}(1+q^2)^{-x/2}\left((1+q^2)^{-x/2}1-q^x-1\right)$.

Note that $g_1(q,x)\,,g_2(q,x)\propto q^{3-x}$ when $q\to 0$, and $g_2(q,x)\propto q^{1-2x}$ when $q\to \infty$, we find that $I(x)$ is convergent for $1<x<4$. Using Mathematica, we get the analytical result:
\begin{equation}\label{eq:Ishort}
    I(x)=\frac{2}{x-2}-\frac{\sqrt{\pi
   } 2^{x-3} \left(\sec
   \left(\frac{\pi 
   x}{2}\right)-1\right)
   \Gamma
   \left(1-\frac{x}{2}\right)}
   {\Gamma
   \left(\frac{3}{2}-\frac{x}{
   2}\right)}\,,
   \end{equation}
and plot it in \figref{fig:Ix}(b).
With \eqref{eq:Ishort}, one can easily confirm that $I(3)=0$, and find another solution $x=2$ that corresponds to the thermal equilibrium spectrum. 

The derivative $I'(3)$ then becomes $I'(3)=\int_0^1g'_1(q,3)\rmd q - 2\int_0^{\infty}g'_2(q,3)\rmd q= \pi-4\sqrt{2}$. Finally, one gets the KZ spectrum for the short Bogoliubov waves, as in Eq.~(3) of the main text.

\section{Verifying the assumptions of the Wave Turbulence theory for the GPE simulations}
\vspace{-4 ex}
\begin{figure*}[h]
\centering
\includegraphics[width=0.98\textwidth]{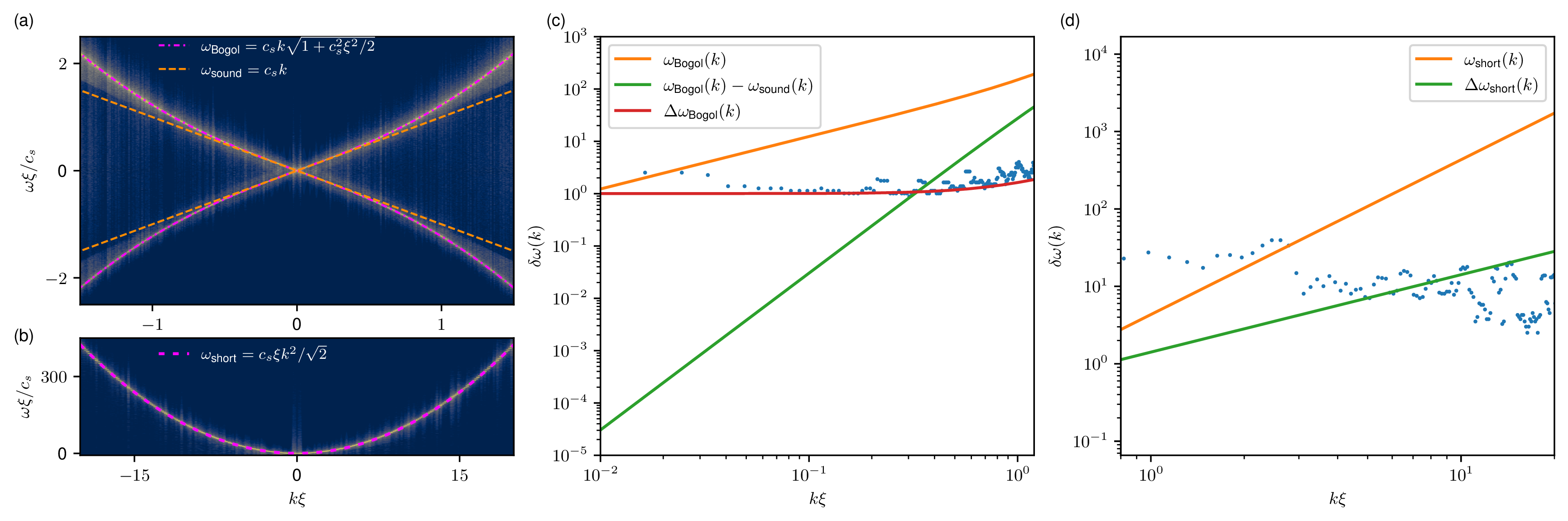}%
\caption{\label{fig:KW}
Normalized spatio-temporal spectral density of $\Psi(\bm{x}, t)$ for the GPE simulations of (a) run 3, and  (b) run 4; 
Frequency broadening $\delta\omega(k)$ (blue points) extracted from the corresponding spatial-temporal density spectra for (c) run 3, and  (d) run 4.}
\vspace{-1 ex}
\end{figure*}

To check if the WTT assumptions apply to the GPE simulations, we compute 
the normalized spatio-temporal spectral density $S(\omega, k) \propto |\hat{\Psi}(k, \omega)|^2$, 
where $\hat{\Psi}(k, \omega)$ is the time and space Fourier transform of $\Psi(\bm{x}, t)$, averaged on the sphere  $|\bm{k}|=k$. We present the spatio-temporal spectra in \figref{fig:KW} (a) for GPE run 3  and (b) for run 4.
\figref{fig:KW} (a) demonstrates that for run 3, most of the spectrum is concentrated near the 
 Bogoliubov dispersion $\omega_{\text{Bogol}}(k)=\cs k \sqrt{1+k^2\xi^2/2}$, which coincides with the dispersion relation of acoustic waves  $\omega_{\text{sound}}(k)=\cs k$ for $k\xi < 0.5$. For the short-wave simulation run 4,
 \figref{fig:KW} (b) indicates that most waves concentrate around $\omega_{\text{short}}(k)=\cs \xi k^2/\sqrt{2}$. 

The nonlinearity of the systems is estimated by measuring the frequency broadening $\delta\omega(k)$ around $\tilde{\omega}(k)$ such that $S(\omega, k)$ gets its maximum value 
for each fixed $k$.
We plot the frequency broadening of run 3 and run 4 in \figref{fig:KW}(c) and (d), respectively. The WWT theory  requires that: 1) The linear frequency is much bigger than nonlinear frequency broadening,  $\omega_{\text{Bogol}}(k) > \delta\omega(k)$; 2) A stronger requirement for the weakly-dispersive waves \cite{ZLF}: the dispersion correction to non-dispersive frequency, $\omega_{\text{disp}}(k)=\cs k(\sqrt{1+k^2\xi^2/2}-1)$, is greater than $\delta\omega(k)$; 3) To trigger the interactions, 
the nonlinear broadening must be greater than the spacing between the eigenmodes of the periodic box, $\delta\omega(k) > \Delta\omega_{\text{Bogol}}(k)$, where $\Delta \omega_{\text{Bogol}}(k)=\Delta k\,\rmd \omega_{\text{Bogol}}(k)/\rmd k$ and $\Delta k= 2\pi/L$ is the $k$-grid spacing. According to \figref{fig:KW}(c),  $0.2 < k\xi <1$ meets the three conditions and it is also the range where the energy spectrum obtained by GPE run 3 agrees well with the one obtained by WKE run 1, as shown in Fig.~1 of the main text. As noted, \cite{kochurin2024three} obtained ZS spectrum in non-dispersive acoustic waves, which means condition 2) might not be necessary (this complicated issue is far from being settled). 
WWT for short waves requires that $\omega_{\text{short}} (k)> \delta\omega(k) $ and 
$\delta\omega(k) > \Delta\omega_{\text{short}}(k)$, where $\Delta \omega_{\text{short}}(k)=\Delta k\,\rmd \omega_{\text{short}}(k)/\rmd k$.
The $k\xi$-range that satisfies the two constraints of WTT, as shown in \figref{fig:KW}(d), are 2-15 for run 4. This range is consistent with the one where expected KZ spectra are observed in the main text. Details on how the normalized spatio-temporal spectral density and frequency broadening are calculated can be found in \cite{zhu2022testing}.

\vspace{-2 ex}
\end{widetext}

\end{document}